\documentclass[useAMS,usenatbib]{mn2e}
\usepackage{graphicx}
\usepackage{color}
\usepackage{url}
\newcommand{\Msol}{\ensuremath{\mathrm{M_{\odot}}}}

\newcommand{\nm}{\mbox{\ensuremath{\mathrm{~\nm}}}}

\newcommand{\degree}{\ensuremath{\mathrm{^\circ}}}

\newcommand{\rxte}{\emph{RXTE}}
\newcommand{\he}{HEAsoft}
\newcommand{\xspec}{\emph{Xspec}}
\newcommand{\ftool}{\emph{ftool}}
\newcommand{\compps}{{\sc compps}}
\newcommand{\phabs}{{\sc phabs}}
\newcommand{\diskbb}{{\sc diskbb}}
\newcommand{\totalmod}{{\sc const(phabs(compps+gauss)) }} 
 
\newcommand{\change}[1]{{ #1}} 
\newcommand{\newchange}[1]{{ #1}} 
\usepackage{amsmath}
\usepackage{amssymb}
\usepackage{float}

\title{A dichotomy between the hard~state~spectral properties of black~hole and neutron~star X-ray~binaries}
\author[M. J. Burke, M. Gilfanov  \& R. Sunyaev]{M. J. Burke$^{1}$\thanks{E-mail:
mburke@mpa-garching.mpg.de (MJB);}, M. Gilfanov$^{1,2,3}$ and R. Sunyaev$^{1,2}$    \\
$^{1}$ Max Planck Institute for Astrophysics, Karl-Schwarzschild-Str. 1, Garching b. Munchen D-85741, Germany\\
$^{2}$ Space Research Institute of Russian Academy of Sciences, Profsoyuznaya 84/32, 117997 Moscow, Russia \\
$^{3}$ Kazan Federal University, Kremlevskaya str.18, 420008 Kazan, Russia}

\begin{document}

\date{Accepted Year Month Day. Received Year Month Day; in original form Year Month Day}

\pagerange{\pageref{firstpage}--\pageref{lastpage}} \pubyear{2015}
\maketitle

\label{firstpage}

\begin{abstract}

We analyse the spectra of black hole (BH) and neutron star (NS) X-ray binaries (XBs) in the hard state using archival RXTE observations. We find that there is a clear dichotomy in the strength of Comptonisation between NS and BH sources, as measured by both the Compton $y-$parameter and amplification factor $A$, with distinct groups of BH and NS XBs separated at $y\sim 0.9$ and $A\sim 3$.   The electron temperature $kT_e$ can occupy a broad range in BH systems, from $kT_e\sim30-200$ keV, whereas for NSs $kT_e$ is peaked at $\sim15-25$ keV, but can extend to higher values.  The difference between BHs and NSs in $y$ implies that $kT_e$ is higher at a given optical depth for BH XBs.  Our results also imply that for NS systems the accreting material loses $\sim 1/2-2/3$ of its energy through Comptonisation in the corona. The remaining energy is released on the surface of the neutron star, making it a powerful source of soft radiation, which alters the properties of the Comptonising corona.   Finally, we find evidence at the $\sim2.4\sigma$ confidence level that Comptonisation parameters may be correlated with the neutron star spin, whereas no correlation with the BH spin is found.  Our results highlight a further observational distinction between BH and NS XBs that is a consequence of NSs possessing a physical surface.

\end{abstract}


\section{Introduction}
X-ray binaries (XBs) are observed in a rich variety of states and phenomenologies \citep[e.g.][]{2006ARA&A..44...49R,2010LNP...794...53B,2010LNP...794...17G}. However, from a broad perspective their behaviour can be reduced to a discussion of two spectral states. These are comprised of a soft state, where the spectra peak below 10 keV and are well-described as optically thick emission from a geometrically thin disc \citep{1973A&A....24..337S}, and a hard state where the spectra follow a roughly power law shape to tens or hundreds of keV \citep[e.g.][]{1991SvAL...17..409S,1996ARA&A..34..607T,2005MNRAS.362.1435I, 2010LNP...794...17G}.  

The precise emission geometry of the hard state remains unknown, however, the observed photon index $\Gamma$ is consistent with the bulk of the emission arising from the unsaturated Comptonisation of seed photons by a hot electron cloud of temperature $kT_e$ and optical depth $\tau$ \citep{1980A&A....86..121S}.  The interplay between these two properties can be accounted for by describing the spectral shape in terms of the Compton $y$-parameter 

\begin{equation}
\label{eq:comp}
y= \frac{4kT}{mc^2}Max(\tau,\tau^2) ,
\end{equation}
which is a measure of the average change in energy a population of photons will experience while travelling through a finite medium.  

Comptonisation is a cooling process, where hot electrons impart energy through scattering with lower energy photons, which means that the supply and temperature of the seed photons will affect the properties of the Comptonising media.  \cite{1989ESASP.296..627S} showed that if the seed photons originate in the accretion disc and close to the Comptonising cloud then it is necessary to account for photon feedback, where Comptonised photons are subsequently absorbed by the disc, increasing the temperature and numerosity of seed photons going into the cloud, causing more Comptonisation and a greater cooling of the Comptonising media.    In the case of a uniform thermal electron cloud blanketing the disc, this will result in softer (and more rapidly softening) spectra than is typically observed from BH XBs \citep{1993ApJ...413..507H}, thus strongly disfavouring such a scenario \citep{1995ASIC..450..331G, 2001MNRAS.321..759C}.  The hard state geometry is therefore subject to two important constraints; the absence of the the soft state emission consistent with an optically thick accretion disc (with $kT_i\sim 1$~keV), and the positioning of the two emission regions (seed photon source and Comptonising electron cloud) such that the effect of photon feedback is reduced.

Some fraction of the Comptonised photons will be intercepted by the disc and `reflected' into the line-of-sight by further Compton scattering and florescence.     These processes give rise to a complex spectral component that typically peaks at energies of $\approx30-50$ keV \citep[the so-called `Compton hump', see][Fig. 10a,c]{1980A&A....86..121S} and adds further fluorescent emission to the spectrum.  The reflection component is present in both BH XBs \citep[due to interaction of X-rays with the accretion disc and the surface of the normal star,][]{1974A&A....31..249B} and active galactic nuclei \citep{1994MNRAS.268..405N}, and complicates attempts to understand the dominant Comptonised emission owing to proximity of the Compton hump to the plausible location of the high-energy turn-off (corresponding to $\approx 3kT_e$).  Typical attempts to take reflection into account involve adding sophisticated reflection components (such as the \xspec~models \emph{pexrav} \citep{1995MNRAS.273..837M} or \emph{reflionx} \citep{2005MNRAS.358..211R}) to unsophisticated models of Comptonisation, such as a cut-off power law.  This inaccurate representation of the Comptonisation means that interesting properties such as the electron temperature or Compton $y$-parameter are then inferred from the best fit of a cut-off power law to the data, rather than a direct treatment of these quantities.  In addition to this, \cite{2005MNRAS.362.1435I} showed that the strength of the reflected component $R$ (defined in terms of the solid angle $\Omega$ subtended by the portion of the disc in line-of-sight to the corona) is systematically over-estimated by such treatment.   It has been known for some time that $R$ correlates strongly with photon index $\Gamma$ \citep{1999A&A...352..182G,2001A&A...380..520R,2005MNRAS.362.1435I},  increasing as spectra become softer (implying a decreasing $y$-parameter).  Similar behaviour has been observed for AGN spectra \citep{1999MNRAS.303L..11Z,  2000sgwa.work..114G,2003MNRAS.342..355Z}. This relationship clearly favours the accretion disc as the source of the seed photons for Comptonisation, as the radiation being intercepted by the disc is increasing in tandem with the seed photon flux incident on the electron cloud.     However, there is increasing evidence that at lower luminosities synchrotron photons from the magnetised corona may contribute a significant, perhaps dominant, population of seed photons, motivating the creation of so-called hybrid models.  Such models typically comprise both a low-energy thermal population and an additional high-energy non-thermal population of electrons \citep[see e.g.][]{1999ASPC..161..375C,2001MNRAS.321..549M,2013MNRAS.430..209D}.

Typical attempts to understand the geometry of the hard state either consider a truncated disc with an electron cloud close to the compact object \citep[e.g.][]{1997MNRAS.292L..21P}, where the innermost portions of the disc have either evaporated to form an optically thin, hot flow \citep{1994A&A...288..175M} or are absent altogether, with the hard X-ray emission stemming from the base of the jet \citep{2001A&A...372L..25M,2005ApJ...635.1203M}, which is always concurrent with the hard state \citep{2004MNRAS.355.1105F}.  Other plausible geometries invoke the presence of non-uniform or dynamic coronae \citep{1999ApJ...510L.123B}. Possible hard state geometries are discussed in detail by \cite{2007A&ARv..15....1D} and \cite{2010LNP...794...17G}.  

A majority of NS XBs also exhibit soft and hard states~\citep{1989A&A...225...79H}, but differ observationally from BH XBs \citep[e.g.][]{2010LNP...794...17G} in that the hard state is always at a relatively low luminosity in the range $10^{36}-10^{37}~{\rm erg~cm^{-2}~s^{-1}}$ (${\rm 0.01-0.1L_{Edd}}$), what is often referred to as the `island state'.  The majority of observational differences arise due to the presence of a physical surface.  The difference in angular momentum between the innermost portions of the disc and the surface leads to the formation of a boundary layer, an ever-present \citep{1988AdSpR...8..135S} portion of the flow where the kinetic energy of the accreting material is surrendered \citep{2000AstL...26..699S}.  In addition, material will accumulate on the NS surface until the critical density for thermonuclear ignition is reached, which leads to powerful explosions that are observed as type-I X-ray bursts \citep[see][for review]{2006csxs.book..113S}.   

\cite{2003A&A...410..217G} used Fourier-resolved spectroscopy to show that the boundary layer has a similar spectrum across a handful of sources, and can be approximated by a Wien spectrum of characteristic temperature $kT\sim 2.4$~keV.  The emission from the boundary layer, or even the NS surface itself, should provide additional seed photons for Comptonisation in the hard state, affecting the properties of the Comptonising media \citep{1989ESASP.296..627S}.  Indeed, it is observed that the hard state of NSs is generally softer than that of BHs \citep{1991SvAL...17..409S,1993ApJ...418..844G,1995ApJ...443..341C,2003MNRAS.342.1041D}, which could be attributed to the  Comptonising corona being cooler in such systems. We emphasise that when discussing the hard state in this paper we refer specifically to the traditional `low/hard' state, where the source is at comparatively low luminosity ($\lesssim 10^{37}~{\rm erg~s^{-1}}$, ${\rm 0.1L_{Edd}}$ in the case of NSs) and not the states that can occur at higher luminosities when a strongly Comptonised component is also present in the spectrum. 

Distinguishing BH from NS in an XB requires dynamical mass measurements obtained by optical or IR spectroscopy of transient sources during quiescence, when the optical emission is dominated by the companion \citep[see][for review]{2006csxs.book..215C}.  However,  our ability to firmly identify BHs is limited by the necessity for a source to be transient and for the secondary to be observable with optical instrumentation.  Such limitations mean that formally discerning BH from NS in extragalactic studies is practically impossible. Therefore, there is an increasing list of behaviours that can be used to identify an XB as a black hole `candidate' (BHC), or definitively as a NS. Most strikingly, NS LMXBs display temporal behaviour that is not observed from BH LMXBs, such as type-I X-ray bursts and, when in the hard state, a significant amount of noise in their X-ray variability for frequencies greater than $\sim500$~Hz \citep{2000A&A...358..617S}.  While BHs and NSs exhibit similar X-ray states, there are some important state-dependent spectral differences, as BH sources should be softer during the so-called soft or thermal dominant state when the strongest source of emission is from the disk.  This emission is characterised by the temperature of innermost portion of the disk, its position determined by the ISCO of the compact object, which will be larger for BHs than for NSs (for which the minimum inner disc radius can be also set by the radius of the NS).  In this paper we propose another spectral technique for distinguishing the nature of the compact object in XBs based on the Comptonisation properties of their coronae in the hard state.

The paper is organised as follows.  In \S~\ref{sec:red} we assemble a sample of BH and NS XBs for which RXTE observations  of the hard states are available and outline the subsequent reduction process.  We also outline the reduction of a handful of archival, simultaneous XMM-Newton observations (\S~\ref{subsec:xxmr}).  In \S~\ref{sec:anal} we analyse RXTE spectra over a broad energy range ($3-200$ keV) with the sophisticated Comptonisation model of \cite{1996ApJ...470..249P}, allowing us to separate out the properties of the seed photon population, reprocessing materials and the Comptonising region, then present our results in \S~\ref{subsec:specanal}.  Discussion in \S~\ref{subsec:yparm} demonstrates that NSs have a systemically lower Compton $y$-parameter and we posit that this is a consequence of the physical surface providing additional seed photons for Comptonisation, producing different physical properties of the Comptonising media. The diagnostic potential of the observed differences in hard state spectra is considered, both in the context of Galactic and Extragalactic X-ray binary studies.   We then investigate surprisingly high seed photon temperatures (\S~\ref{subsec:seed}), and use simultaneous XMM observations to show that these are systematically over-estimated by fitting to the \rxte~bandpass and establish that this does not alter significantly the principal conclusions of this work.     We summarise our conclusions in \S~\ref{sec:conc}.

\section{Data Reduction}
\label{sec:red}
\subsection{RXTE Data}
We first identified a selection of known BH and NS XBs \citep{2003A&A...404..301R} where repeated RXTE observations exist in the HEASARC archive\footnote{\url{https://heasarc.gsfc.nasa.gov/db-perl/W3Browse/w3browse.pl}}.  In the case of NSs, our requirement that the source exhibits the classic hard (aka `island') state meant that no Z-track sources were considered.   As a preliminary step we constructed lightcurves for each source, covering the whole \rxte~mission, in order to identify periods when each source was in the hard state.  For each sub-observation we retrieved the PCA standard2 data products, housekeeping data and filter file.  These files allow us to produce deadtime corrected, background uncorrected lightcurves and hardness-intensity diagrams (HIDs) without having to download the entire \rxte~ archive of observations for each source.  For this exercise we use data exclusively from PCU2, which was always turned on during observations.  Using \he~6.15 we extract 16s-binned lightcurves from three energy ranges; $4.00-6.00$ keV, $6.00-7.50$  keV and $7.50-18.50$. In order to compensate for long-term variation in the gain of the detector we converted to absolute channels, which depend on the mission epoch\footnote{\url{http://heasarc.gsfc.nasa.gov/docs/xte/e-c_table.html}}.  Using the intensities $I$ for each band we calculate a hardness ratio $H=I_{7.50-18.50}/I_{6.00-7.50}$ for each point in the lightcurve. This hardness ratio is similar to that used by \cite{2007ApJ...667.1073L}, who showed that for typical NS XBs the hard and soft states are fully distinguishable by comparing the intensity in the PCA above and below $\approx8$~keV.  Based on inspection of the hardness-intensity diagrams, we chose sub-observations for analysis where $H\ge2.0$ for the majority of time bins.   We apply an additional constraint by considering data only where $I_{4.0-18.0}>50~{\rm counts~s^{-1}~PCU^{-1}}$, in the interest of achieving good statistics for spectral fitting as well as consistent handling of the background for all sources.  Where possible we include data from multiple outbursts of a given source.  A list of the sources studied in this work is presented in table~\ref{tab:nh} together with the assumed distance, BH mass and equivalent Hydrogen column density used in our analyses.  

The hard state datasets used in this work are presented in table~\ref{tab:sources}.  After downloading the full dataset for each sub-observation, we produced full 64s-binned lightcurves using both the left- and right-anode chains from the first Xenon layer PCU2 data that were both deadtime and background corrected (using the bright background model). In the interests of good calibration we chose not to include PCU0 or PCU1 data obtained after the failure of their respective propane layers in 2000 and 2006.  Further, we filtered these data to exclude periods when the pointing was greater than 0.02 degrees off-target, when the elevation was less than 10 degrees, in the period immediately preceding the passage of the satellite through the South Atlantic Anomaly (SAA), and periods of 600s after PCA breakdown events.  We determined the optimal PCA configuration for each observation, so as to maximise the number of counts in the eventual spectra, and produced lightcurves and hardness ratios using these configurations.  Good time interval (GTI) files were produced based on these lightcurves, covering continuous portions of the observation and also ignoring any short-term events such as X-ray bursts.  These GTIs have typical durations of $\sim1-3$~ks.       

We extract source and background PCA spectra for each GTI, again using both left and right anodes.  Both source and background are subsequently deadtime corrected and response files generated.  The PCA spectra are then rebinned such that there are a minimum of 40 counts per bin, which enables the use of the $\chi^2$ statistic in spectral fitting.  As is standard practice, we add a $0.5\%$ systematic uncertainty to each bin.   

\begin{table}
  \begin{tabular}{lcccr}
Source & ${\rm N_H}$ & Distance & Mass & Ref. \\
 & $\rm{10^{22}~cm^{-2}}$ & kpc & \Msol &\\
\hline
GS~1354-64 & 0.73 &   $26\pm1.0 $ & $>7.6$& A \\
GRO~J1654-40 & 0.53 &  $3.2\pm0.2 $ & $6.6\pm0.5$ & B,M \\ 
GX~339-4 & 0.37  &  $10\pm4 $ &  $>6.0$ & C,L \\ 
XTE~J1550-564 & 1.01  &   $5.3\pm2.3 $ & $7.8-15.6$ & D,N \\ 
XTE~J1118+480 & 0.013 & $1.72\pm0.1 $ & $6.9-8.2$ & E,O \\ 
Cyg~X-1 & 0.721 &  $ 1.86^{+0.11}_{-0.12}$ & $14.8\pm1.0$ & F,P \\
4U~1543-47 & 0.35 &  $ 7.5\pm 0.5$ & $2.7-7.5$ & K \\ 
4U~1636-536 &  0.27  &  $  6.0\pm0.5 $ & & G \\ 
4U~1705-44 & 0.67 &  $ 8.4^\pm1.2$ & & J \\ 
4U~1728-33 & 1.24  & $4.6\pm 0.2 $ &  & I \\ 
Aql~X-1 & 0.28  & $5.2 \pm 0.8 $ & & J \\ 
4U~1608-52 & 1.81  & $3.3\pm 0.5$ & & J \\
  \end{tabular}
  \caption{Source properties.  We make use of equivalent Hydrogen column densities inferred from 21 cm emission \citep{2005A&A...440..775K}, plus the best measurements of distance and compact object mass for each source. References correspond as A:~\protect\cite{2009ApJS..181..238C}, B:~\protect\cite{1995Natur.375..464H}, C:~\protect\cite{2004ApJ...609..317H},D:~\protect\cite{2002ApJ...568..845O}, E:~\protect\cite{2006ApJ...642..438G}, F:~\protect\cite{2011ApJ...742...83R}, G:~\protect\cite{2006ApJ...639.1033G}, H:~\protect\cite{2014SSRv..183..223C}, I:~\protect\cite{2003ApJ...590..999G}, J:~\protect\cite{2004MNRAS.354..355J}, K:~\protect\cite{1998ApJ...499..375O}, L:~\protect\cite{2008MNRAS.385.2205M}, M:~\protect\cite{2003MNRAS.339.1031S},N:~\protect\cite{2011ApJ...730...75O}, O:~\protect\cite{2013AJ....145...21K}, P:~\protect\cite{2011ApJ...742...84O}.  For further discussion on BH mass estimates see \protect\cite{2014SSRv..183..223C} \label{table:nh}.}

  \label{tab:nh}
\end{table}

HEXTE source and background spectra were extracted over the same GTI files using the \ftool~\emph{HXTLCURV}, and the correct response files generated.  We use both cluster A and cluster B data prior to 2004, when cluster A ceased to move between on- and off-target pointings, and use only cluster B data for the period between 2004 and December 2010.  This process also produced 64s-binned background-subtracted lightcurves, which were used in the creation of good time intervals.

\begin{table*}
  \begin{tabular}{lccccc}
    {\large Source Name} & {\large Observations} & PCUs &  Exposure (ks) & \\ \hline  
{\bf Black Holes} & & & {\bf PCA} & {\bf HEXTE-B} & {\bf HEXTE-A}  \\     
GS~1354-64 & 20431-01-03-00 & 0,1,2,3,4 & 26.787 & 1.666 & 1.67  \\
GS~1354-64 & 20431-01-04-00 & 0,1,2,3,4 & 13.58 & 0.774 & 0.792  \\
GS~1354-64 & 20431-01-05-00 & 0,1,2,3,4 & 12.493 & 0.776 & 0.776  \\
GS~1354-64 & 20431-01-05-00 & 0,1,2,3,4 & 15.6 & 0.969 & 0.999  \\
4U~1543-47 & 70124-02-06-00 & 2,3 & 1.692 & 0.293 & 0.293  \\
4U~1550-564 & 30188-06-01-00 & 0,1,2,3,4 & 2.014 & 0.129 & 0.126  \\
4U~1550-564 & 30188-06-01-01 & 0,1,2,3,4 & 3.249 & 0.237 & 0.259  \\
4U~1550-564 &  30188-06-01-02 & 0,1,2,3,4 & 5.905  & 0.362 & 0.368  \\
4U~1550-564 & 50134-02-01-00 & 0,1,2,3,4 & 3.872 & 0.255 & 0.255  \\
4U~1550-564 & 50135-01-03-00 & 2,3,4 & 6.781 & 0.674 & 0.662  \\
4U~1550-564 & 50135-01-05-00 & 2,3,4 & 3.594 & 0.403 & 0.391  \\
4U~1550-564 & 50135-01-06-00 & 2,3 & 2.743 & 0.475 & 0.476  \\
4U~1550-564 & 50137-02-07-00 & 2,3,4 & 1.404 & 0.18 & 0.158  \\
Cyg~X-1 & 20173-01-01-00 & 0,1,2,3,4 & 9.639 & 0.621 & 0.604  \\
Cyg~X-1 & 20173-01-01-00 & 0,1,2,3,4 & 14.492 & 0.922 & 0.909  \\
Cyg~X-1 & 20173-01-01-00 & 0,1,2,3,4 & 16.237 & 1.025 & 1.003  \\
Cyg~X-1 & 20173-01-01-00 & 0,1,2,3,4 & 16.936 & 1.029 & 1.022  \\
Cyg~X-1 & 20173-01-02-00 & 0,1,2,3,4 & 12.704 & 0.789 & 0.786  \\
Cyg~X-1 & 20173-01-02-00 & 0,1,2,3,4 & 15.66 & 0.969 & 0.945  \\
Cyg~X-1 & 20173-01-02-00 & 0,1,2,3,4 & 16.044 & 0.969 & 0.974  \\
Cyg~X-1 & 80110-01-43-00 & 2,3 & 6.348 & 0.989 &    \\
Cyg~X-1 & 80110-01-44-00 & 2,3 & 3.318 & 0.584 &    \\
Cyg~X-1 & 80110-01-45-00 & 2,3 & 3.38 & 0.621 &    \\
Cyg~X-1 & 94108-01-01-00 & 2 & 2.982 & 1.028 &    \\
GX~339-4 & 92035-01-01-01 & 2 & 3.049 & 1.139 &    \\
GX~339-4 & 92035-01-01-02 & 2 & 3.581 & 1.243 &    \\
GX~339-4 & 92035-01-02-04 & 2 & 3.071 & 0.962 &    \\
GX~339-4 & 92035-01-02-04 & 2 & 2.99 & 0.999 &    \\
GX~339-4 & 90118-01-06-00 & 2 & 1.520 & 0.523 & 0.522   \\
XTE~J1118+480 & 50133-01-01-00 & 2,3,4 & 6.757 & 0.703 & 0.733  \\
XTE~J1118+480 & 50133-01-02-00 & 2,3,4 & 7.98 & 0.859 & 0.868  \\
XTE~J1118+480 & 50133-01-02-01 & 2,3,4 & 6.636 & 0.757 & 0.767  \\
XTE~J1118+480 & 50133-01-02-01 & 2,3,4 & 7.439 & 0.869 & 0.864  \\
XTE~J1118+480 & 50133-01-02-01 & 2,3,4 & 6.741 & 0.79 & 0.788  \\
XTE~J1118+480 & 50133-01-03-00 & 2,3,4 & 9.478 & 1 & 1.026  \\
XTE~J1118+480 & 50133-01-03-00 & 2,3,4 & 9.608 & 0.985 & 1.03  \\
GRO~J1654-40 & 50133-01-03-00 & 2,3 & 3.477 & 0.317 &    \\
GRO~J1654-40 & 91702-01-01-03 & 2,3 & 3.273 & 0.53 &    \\
GRO~J1654-40 & 91702-01-01-04 & 2,3 & 5.697 & 0.92 &    \\
GRO~J1654-40 & 91702-01-01-05 & 2,3 & 1.908 & 0.359 &    \\
{\bf Neutron Stars} & & & & & \\4U~1608-52 & 60052-03-01-06 & 2,3,4 & 10.451 & 1.265 & 1.274  \\
4U~1608-52 & 60052-03-02-02 & 2,3,4 & 13.786 & 1.487 & 1.501  \\
4U~1608-52 & 60052-03-02-02 & 2,3,4 & 13.119 & 1.384 & 1.41  \\
4U~1608-52 & 60052-03-02-04 & 2,3,4 & 9.422 & 1.13 & 1.136  \\
4U~1608-52 & 60052-03-02-06 & 2,3,4 & 9.378 & 1.136 & 1.136  \\
4U~1636-536 & 92023-02-08-00 & 2 & 1.949 & 0.666 &    \\
4U~1636-536 & 92023-02-09-00 & 2 & 1.469 & 0.464 &    \\
4U~1636-536 & 92023-02-10-00 & 2 & 1.4 & 0.411 &    \\
4U~1636-536 & 92023-02-11-00 & 2 & 2.023 & 0.696 &    \\
4U~1636-536 & 92023-02-12-00 & 2 & 0.801 & 0.294 &    \\
4U~1636-536 & 94310-01-04-00 & 2 & 2.322 & 0.708 & \\
4U~1705-44 & 20073-04-01-00 & 0,1,2,3,4 & 12.255 & 0.677 & 0.678  \\
4U~1705-44 & 20073-04-01-00 & 0,1,2,3,4 & 13.974 & 0.804 & 0.804  \\
4U~1728-34 & 92023-03-47-00 & 2 & 1.958 & 0.614 &    \\
4U~1728-34 & 92023-03-49-00 & 2 & 2.11 & 0.733 &    \\
Aql~X-1 & 50049-01-04-01 & 1,2,3 & 6.049 & 0.749 & 0.763  \\
Aql~X-1 & 91414-01-07-03 & 2,3,4 & 5.934 & 0.722 &    \\
Aql~X-1 & 91414-01-08-05 & 2,3,4 & 3.703 & 0.433 &    \\
Aql~X-1 & 91414-01-08-07 & 2,3,4 & 5.834 & 0.623 &    \\
  \end{tabular}
  \caption{Sources and observations, with the PCA configuration and the exposure of each spectrum by detector.  }
  \label{tab:sources}
\end{table*}

\subsection{XMM Data}
\label{subsec:xxmr}

To explore the effects of bandpass on our spectral fitting results, we identified three simultaneous XMM observations where three of our sources were in the hard state. We chose to reduce only EPIC-pn data obtained in timing mode to avoid the controversial arguments surrounding the data reduction of bright point sources observed in imaging mode \citep{2010MNRAS.407.2287D,2010ApJ...724.1441M}.  See table~\ref{tab:xmmsource} for a list of XMM observations used, the exposure of the resulting spectrum and the corresponding simultaneous RXTE observation ID.

Each dataset was reduced using SAS release 14.0.0 \citep{2014ascl.soft04004S}. Based on inspection of the $\geq 12$~keV lightcurve, background flares were identified and filtered out into a new event file.  The spectra were extracted from all RAWY and from 5 RAWX either side of the three central columns, which were excluded in the interests of reducing the effect of pile-up. Pile-up was found to be negligible in the resultant spectra using the tool \emph{epatplot}.  The high number of counts present in the wings of the point-spread function mean that any attempt to extract a background spectrum from a supposed `source-free' region will be impossible, and most-likely result in the over-subtraction of background during spectral fitting, skewing the spectral shape \citep[see discussion in][]{2010MNRAS.407.2287D}.  As is now standard practice we chose to assume the contribution from the background spectrum will be negligible in comparison with the source spectra. We applied a grouping of a minimum of 50 counts per spectral bin, and added a $1\%$ systematic uncertainty.  These spectra are analysed and discussed in section~\ref{subsec:seed}.

\section{Analysis}
\label{sec:anal}
Spectral fitting was carried out for each source using an absorbed Comptonisation model with an additional Gaussian component to model the fluorescent Fe emission.  Using the combined PCA and HEXTE spectra, we were able to fit over the $3.0-20.0$ keV and $20.0-200.0$ keV ranges, respectively, for all spectra with the exception of those from P20431 (GS 1354-64), when we used HEXTE spectra above 30 keV owing to calibration uncertainties beneath this energy.  We chose to employ the \xspec~model \compps~ \citep{1996ApJ...470..249P}, which models the Comptonised emission and its reflection by material in the accretion disk.  For \compps~ we leave as free parameters the electron temperature $T_e$, seed photon temperature of a multi-colour disc $T_{bb}$, Compton $y$-parameter, relative reflection factor $R(=\Omega/2\pi)$ and the normalisation.  We fix the relevant parameters such that the electron distribution is Maxwellian, and assume a spherical geometry ($geom=0$) for the treatment of radiative transfer and photon escape probability and a binary inclination of $45\degree$.  All other parameters remain at their default values.  We model the absorption column using \phabs, with $N_H$ fixed at the Galactic value (table~\ref{table:nh}) as reported from the 21~cm survey carried out by \cite{2005A&A...440..775K}.  A multiplicative constant is included in the model so as to account for the difference in calibration between PCA and HEXTE.  Ultimately the spectral model used in \xspec~can be described as \totalmod.

\begin{table}
  \begin{tabular}{llcc}
    {\large Source} & {\large ObsId} & Sim. RXTE & Exp. (ks) \\ \hline     
4U~1636-536 & 0606070401 & 94310-01-04-00 & 25.553 \\
Cyg~X-1 & 0605610401 & 94108-01-01-00 & 19.754 \\
GX~339-4 & 0204730201 & 90118-01-06-00 & 80.022 \\
\hline
  \end{tabular}
  \caption{Archival XMM datasets of XBs in the hard state observed simultaneously RXTE. We list the observation index, simultaneous RXTE observation, and the effective exposure of the final spectra used in spectral fitting (see \S~\ref{subsec:seed}).}
  \label{tab:xmmsource}
\end{table}

Before fitting the spectra we attempted to constrain the calibration constant $C$ by reducing contemporaneous Crab observations for each dataset when the right combination of PCUs were active.  We found while subsequently fitting XBs that the calibration constant recovered from fitting was consistent with that found from fitting the Crab spectra with an absorbed power law, with $C$ in the range $(0.7-0.9) \rm{pcu^{-1}}$.  

We chose to limit the available parameter space to explore, based on reasonable physical criteria.  We restricted the peak energy of the Gaussian to a range of $5.5-7.1$ keV, allowing for the moderate resolution of RXTE in the vicinity of the relativistically broadened Fe K$\alpha$ line, where crudely modelling the excess emission with a Gaussian could conceivably lead to a recovered $E_{peak}$ in the range described.  We also restrict the Gaussian width $\sigma < 2.0$ keV.   We restrict the electron temperature of the plasma $kT_e$ to values above 10 keV, the minimum value for which the numerical method used by the model can be expected to produce reasonable results.  Finally, we constrain the seed photon temperature $kT_{bb}$ to negative values, which forces \xspec~to model the seed photon spectra as emission from a conventional disc blackbody \citep[i.e.][]{1973A&A....24..337S}.

\subsection{Mitigating contamination from Galactic ridge emission}
\label{subsec:GR}
The majority of sources within our sample lie within 10 degrees of the Galactic plane. The low spatial resolution of RXTE ($\approx 1\degree$) means that the spectra of sources in our sample are susceptible to contamination from the Galactic ridge (GR).  The $3.0-20.0$ keV  GR flux can be as high as $6\times10^{-11}~{\rm erg~cm^{-2}}~s^{-1}$~\citep{2003A&A...410..865R}, or $\approx 5{\rm~counts~s^{-1}~PCU^{-1}}$ in RXTE PCA terms.  For many of the sources in our sample, their Galactic longitudes are large enough and/or their count rates are so high ($>500{\rm~counts~s^{-1}~PCU^{-1}}$) as to render any effect on the spectral shape negligible, and we do not take GR emission into account for modelling their spectra.  However, some spectra were observed at count rates of the order $50-300{\rm~counts~s^{-1}~PCU^{-1}}$, and for sources close to the Galactic plane we take further consideration. 

To attempt to correct for the effect of GR emission we add an additional model consisting of a simple power law ($\Gamma=2.15$) and Gaussian line ($E_{line}=6.6$~keV, $\sigma=0.0$~keV with a normalisation tied to that of the power law such as to produce an equivalent width $\approx0.8$~keV), as found by \cite{2003A&A...410..865R}.  We made use of the \xspec~\emph{cflux} component to fix the overall flux contribution of the model between $3.0-20.0$~keV to a value calculated using the NIR-GR X-ray flux relation,
\begin{equation}
 F_{3.0-20.0 keV} {\rm [erg~cm^{-2}~s^{-1}]} \sim0.26\times 10^{-11} F_{3.5}~{\rm [MJy~sr^{-1}]}
\end{equation}
as found by \cite[][]{2006A&A...452..169R}. To obtain the NIR flux in the vicinity of the source, we measured the $3.5\mu m$ flux at the source position on the COBE-DIRBE map \citep{1996ApJ...464L...1B} and corrected for extinction using the relation $A_{3.5\rm \mu m} =0.058A_V$ \citep{1985ApJ...288..618R}, where $A_V$ was determined from the absorption column $N_H$ as presented by \cite{1990ARA&A..28..215D}.

For the majority of sources the GR emission was found to be $<1\times10^{-11}~{\rm erg~cm^{-2}}~s^{-1}$, however, of particular note is the NS source 4U~$1728-33$, for which we estimate the GR flux to be $\sim6\times10^{-11}~{\rm erg~cm^{-2}}~s^{-1}$ due to its position close to the Galactic centre.

\begin{figure}
\begin{center}
\includegraphics[height=0.48\textwidth]{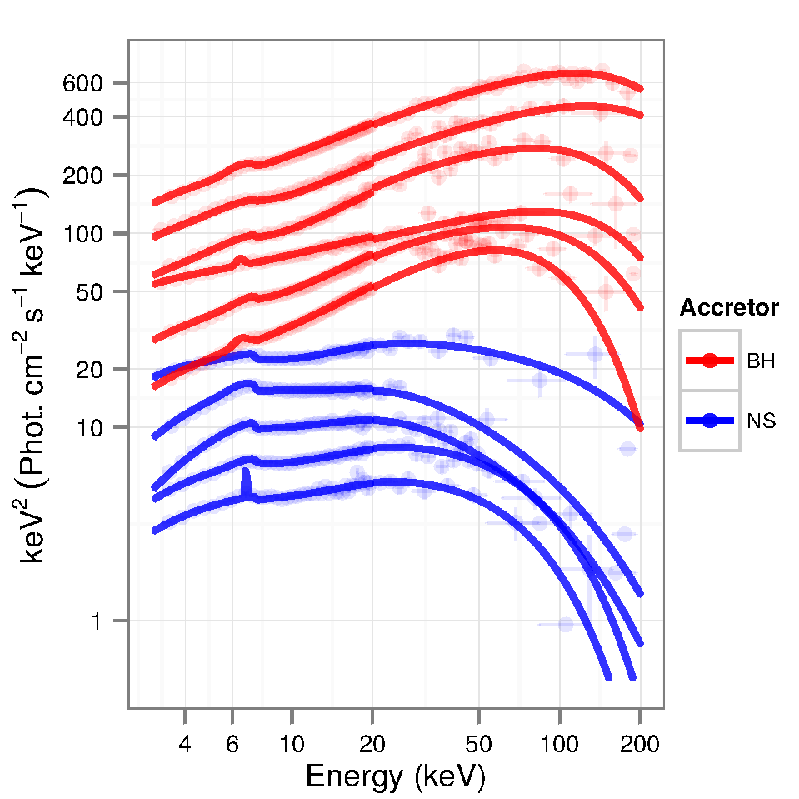}
\caption{Example hard state unfolded RXTE spectra of BH and NS XBs.  The data points were obtained with the {\it Xspec} {\tt plot eeufs} command with the model fixed to a $\Gamma=2.0$ powerlaw, the lines indicate the best-fit \totalmod~model. Spectra are artificially offset in $y-$axis.\label{fig:eeuf}}
\end{center}
\end{figure}

\subsection{\newchange{Parameter Estimation}}
\label{subsec:specanal}

Performing traditional spectral fitting using models containing a \compps~ component with $R$ as a free parameter is computationally expensive; a single spectral fit can take a typical desktop computer many minutes to complete.  A fortiori the calculation of 1D uncertainty intervals using the error command, or even a 2D examination of the parameter space using a steppar grid (a technique that will vary on a source-to-source basis, and should only be used on two parameters at a time).  Therefore, we choose to employ Monte Carlo algorithms to robustly explore the parameter space favoured by our spectra.  Facility for Markov-Chain MC has been integrated into \xspec~for some time, and has proved successful in the past for the calculation of confidence intervals \citep[for example,][]{2013ApJ...766...88B}, however, it is non-trivial to assess how well the chain has converged around the posterior distribution in a consistent way. To bypass this problem we make use of analysis software BXA \citep{2014A&A...564A.125B}, which connects the nested sampling algorithm MultiNest \citep{2009MNRAS.398.1601F} with \xspec. Our methodology followed a two-step approach, primarily fitting the spectra in the nominal way with \xspec, to guarantee that the model can describe the data in a manner acceptable by the $\chi^2$ test, and then using BXA to explore the parameter space.  To find 2D confidence regions we bin over a given 2D plane, rank each bin by the number of MCMC samples it contains and then iteratively sum successive bins of decreasing rank until we have defined an area containing $90\%$ of samples.  In figure~\ref{fig:eeuf} we display examples of the best-fit model and unfolded spectra.

\begin{figure*}
\begin{center}
\hbox{
\includegraphics[width=0.43\textwidth]{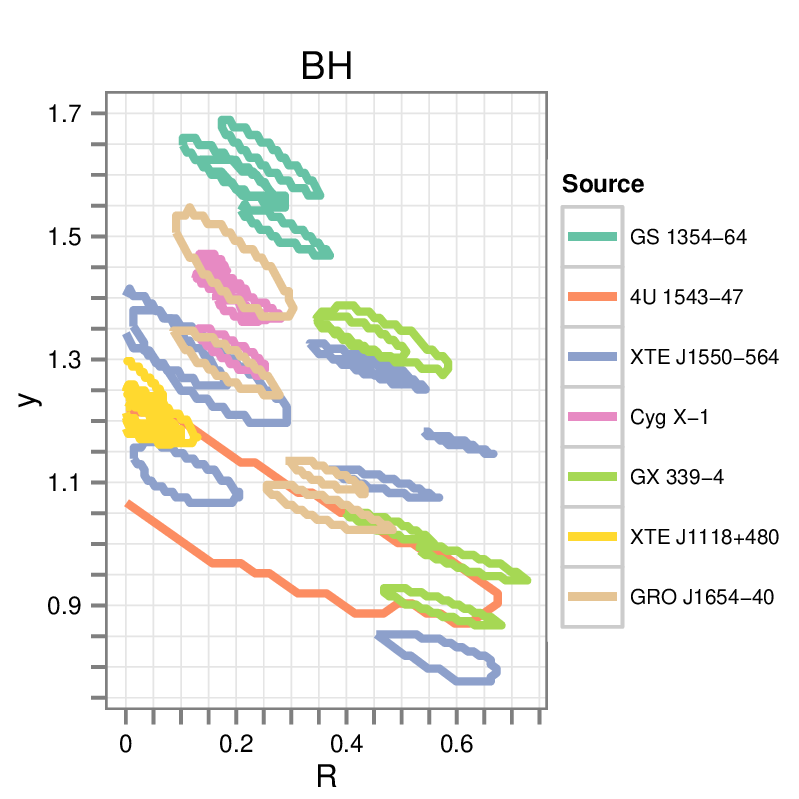} \hspace{1cm}
\includegraphics[width=0.43\textwidth]{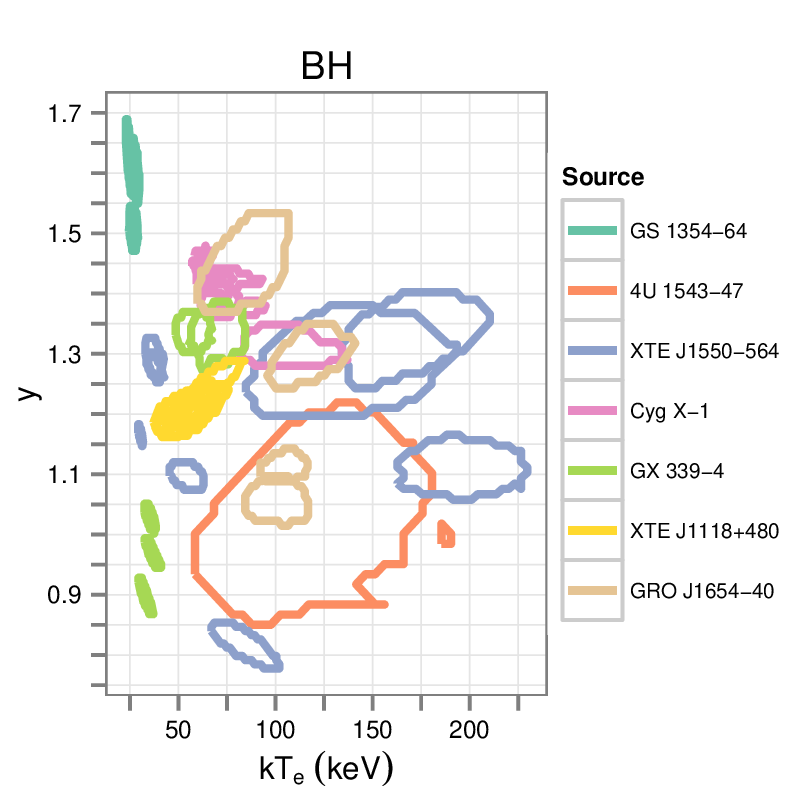} }
\hbox{
\includegraphics[width=0.43\textwidth]{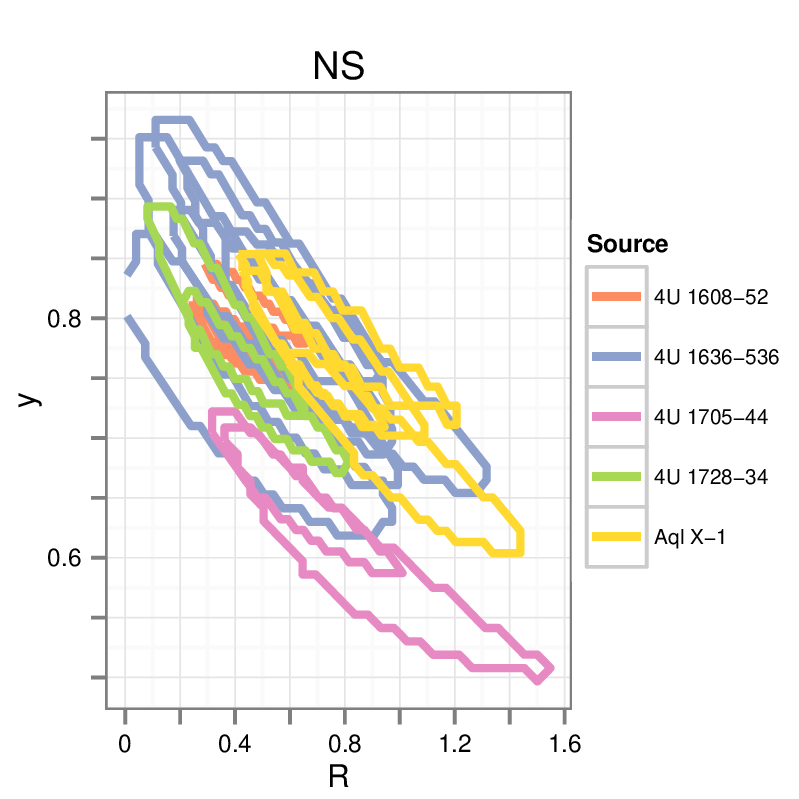} \hspace{1cm}
\includegraphics[width=0.43\textwidth]{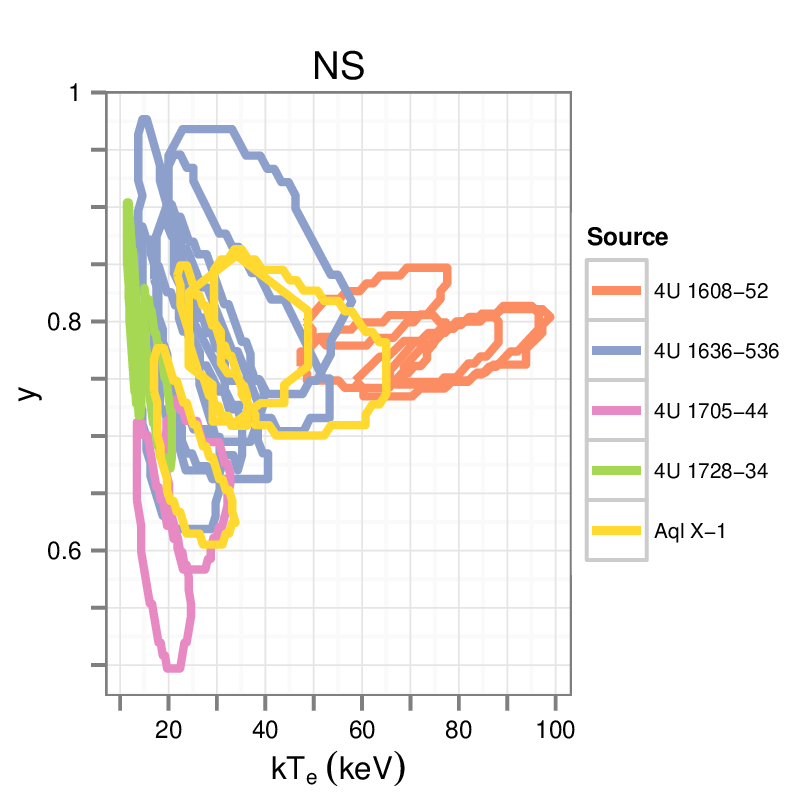}  }
\hbox{
\includegraphics[width=0.41\textwidth]{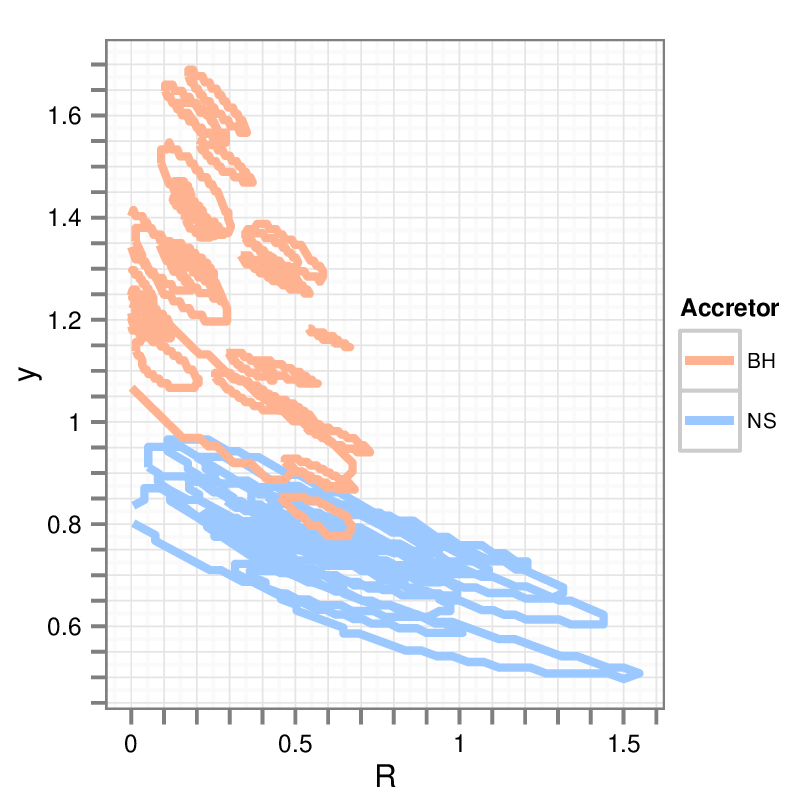} \hspace{1cm}
\includegraphics[width=0.41\textwidth]{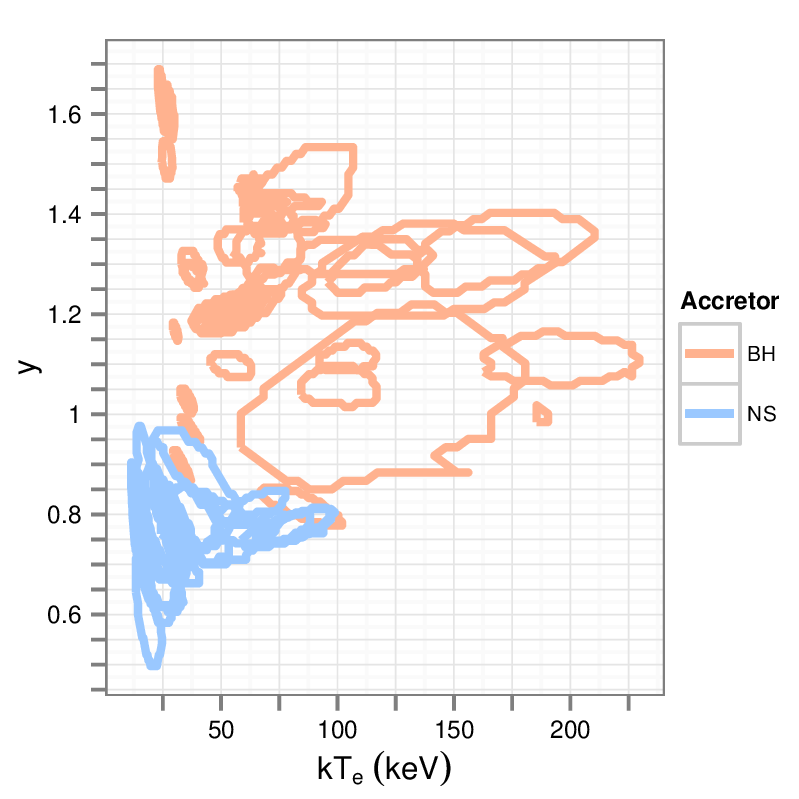}}
\end{center}
\caption{The confidence regions of individual spectra in the $R-y$ (left) and $kT_e-y$ (right) planes for neutron stars and  black holes, where each source is identified by colour, and each contour represents an individual RXTE spectrum (two top rows). The lower panels show the superposition of both results with colour denoting compact object class. \label{fig:NSBH}}
\end{figure*}

We present the 1D confidence intervals of \compps~ parameters in table~\ref{tab:compps}, which we quote at the $90\%$ level throughout this work unless stated otherwise. In the top two rows of figure~\ref{fig:NSBH} we present the $R-y$ and $kT_e-y$ planes for individual BHs and NSs.  In the lower row, we present the two samples together.

\begin{table*}
  \begin{tabular}{lcccccccc}
Source & obsID &  $F_{2-200}$ & $kT_{bb}$~ & $kT_e$~ & R & y & Norm. \\ 
 \hline
GS~1354-64 & P20431 & 5.31  &   $0.94_{-0.14}^{+0.14}$ &   $24.1_{-1.5}^{+1.5}$ &   $0.24_{-0.08}^{+0.08}$ &   $1.64_{-0.05}^{+0.05}$ &   $47_{-19}^{+31}$ \\ 
GS~1354-64 & P20431 & 5.03  &   $0.96_{-0.02}^{+0.09}$ &   $26.3_{-2.0}^{+2.0}$ &   $0.270_{-0.07}^{+0.07}$ &   $1.53_{-0.04}^{+0.04}$ &   $45.5_{-12.0}^{+3.9}$ \\ 
GS~1354-64 & P20431 & 4.73  &   $0.56_{-0.19}^{+0.30}$ &   $25.4_{-3.0}^{+3.3}$ &   $0.18_{-0.09}^{+0.09}$ &   $1.60_{-0.04}^{+0.04}$ &   $258_{-203}^{+934}$ \\ 
GS~1354-64 & P20431 & 4.80  &   $0.83_{-0.10}^{+0.19}$ &   $26.2_{-2.6}^{+2.7}$ &   $0.18_{-0.07}^{+0.07}$ &   $1.62_{-0.05}^{+0.05}$ &   $67_{-34}^{+39}$ \\ 
4U~1543-47 & P70124 & 1.92  &   $0.90_{-0.22}^{+0.22}$ &   $110_{-50}^{+50}$ &   $0.19_{-0.19}^{+0.32}$ &   $1.06_{-0.15}^{+0.15}$ &   $36_{-24}^{+71}$ \\ 
XTE~J1550-564 & P30188 & 26.0  &   $0.71_{-0.02}^{+0.08}$ &   $36.8_{-4.2}^{+4.2}$ &   $0.40_{-0.08}^{+0.09}$ &   $1.30_{-0.03}^{+0.03}$ &   $753_{-246}^{+77}$ \\ 
XTE~J1550-564 & P30188 & 25.7  &   $0.62_{-0.01}^{+0.13}$ &   $38.9_{-3.4}^{+3.4}$ &   $0.450_{-0.07}^{+0.07}$ &   $1.28_{-0.02}^{+0.02}$ &   $1340_{-690}^{+9}$ \\ 
XTE~J1550-564 & P30188 & 29.3  &   $0.42_{-0.02}^{+0.02}$ &   $30.4_{-1.2}^{+1.2}$ &   $0.580_{-0.06}^{+0.06}$ &   $1.17_{-0.02}^{+0.02}$ &   $7240_{-1100}^{+1150}$ \\ 
XTE~J1550-564 & P50134 & 19.4  &   $0.71_{-0.08}^{+0.08}$ &   $77_{-16}^{+18}$ &   $0.56_{-0.09}^{+0.09}$ &   $0.83_{-0.04}^{+0.04}$ &   $1320_{-526}^{+917}$ \\ 
XTE~J1550-564 & P50135 & 8.58  &   $0.83_{-0.03}^{+0.09}$ &   $221_{-45}^{+26}$ &   $0.08_{-0.08}^{+0.08}$ &   $1.10_{-0.05}^{+0.05}$ &   $253_{-93}^{+51}$ \\ 
XTE~J1550-564 & P50135 & 7.45  &   $0.88_{-0.08}^{+0.10}$ &   $201_{-54}^{+26}$ &   $0.02_{-0.02}^{+0.13}$ &   $1.37_{-0.09}^{+0.08}$ &   $127_{-45}^{+59}$ \\ 
XTE~J1550-564 & P50135 & 6.93  &   $0.84_{-0.11}^{+0.11}$ &   $158_{-54}^{+53}$ &   $0.15_{-0.11}^{+0.11}$ &   $1.29_{-0.08}^{+0.08}$ &   $129_{-53}^{+88}$ \\ 
XTE~J1550-564 & P50137 & 20.9  &   $0.66_{-0.02}^{+0.15}$ &   $56.4_{-8.0}^{+7.8}$ &   $0.49_{-0.09}^{+0.09}$ &   $1.09_{-0.03}^{+0.03}$ &   $1080_{-592}^{+158}$ \\ 
Cyg~X-1 & P20173 & 30.0  &   $0.89_{-0.10}^{+0.10}$ &   $63.0_{-5.9}^{+6.9}$ &   $0.15_{-0.03}^{+0.03}$ &   $1.42_{-0.01}^{+0.01}$ &   $353_{-121}^{+184}$ \\ 
Cyg~X-1 & P20173 & 33.8  &   $0.92_{-0.03}^{+0.14}$ &   $62.2_{-3.9}^{+4.0}$ &   $0.16_{-0.02}^{+0.02}$ &   $1.46_{-0.01}^{+0.01}$ &   $331_{-136}^{+39}$ \\ 
Cyg~X-1 & P20173 & 35.6  &   $1.03_{-0.03}^{+0.07}$ &   $63.1_{-2.8}^{+2.8}$ &   $0.160_{-0.02}^{+0.02}$ &   $1.46_{-0.01}^{+0.01}$ &   $235_{-48}^{+23}$ \\ 
Cyg~X-1 & P20173 & 31.7  &   $0.94_{-0.04}^{+0.18}$ &   $60.8_{-3.3}^{+3.3}$ &   $0.16_{-0.02}^{+0.02}$ &   $1.45_{-0.01}^{+0.01}$ &   $295_{-136}^{+46}$ \\ 
Cyg~X-1 & P20173 & 28.5  &   $0.820_{-0.10}^{+0.10}$ &   $70.9_{-12}^{+16}$ &   $0.170_{-0.04}^{+0.04}$ &   $1.43_{-0.02}^{+0.02}$ &   $472_{-188}^{+310}$ \\ 
Cyg~X-1 & P20173 & 29.2  &   $0.79_{-0.01}^{+0.10}$ &   $75.1_{-9.5}^{+9.2}$ &   $0.18_{-0.03}^{+0.02}$ &   $1.42_{-0.01}^{+0.01}$ &   $539_{-199}^{+5}$ \\ 
Cyg~X-1 & P20173 & 28.4  &   $0.880_{-0.01}^{+0.10}$ &   $67.3_{-5.0}^{+5.0}$ &   $0.16_{-0.02}^{+0.02}$ &   $1.42_{-0.01}^{+0.01}$ &   $362_{-123}^{+18}$ \\ 
Cyg~X-1 & P80110 & 48.2  &   $1.01_{-0.10}^{+0.11}$ &   $87.0_{-6.1}^{+6.1}$ &   $0.240_{-0.02}^{+0.02}$ &   $1.37_{-0.01}^{+0.01}$ &   $394_{-125}^{+182}$ \\ 
Cyg~X-1 & P80110 & 31.5  &   $1.08_{-0.11}^{+0.13}$ &   $75.1_{-5.9}^{+5.9}$ &   $0.20_{-0.03}^{+0.03}$ &   $1.39_{-0.02}^{+0.02}$ &   $191_{-64}^{+92}$ \\ 
Cyg~X-1 & P80110 & 32.9  &   $1.27_{-0.09}^{+0.09}$ &   $69.3_{-5.9}^{+5.5}$ &   $0.24_{-0.03}^{+0.03}$ &   $1.39_{-0.02}^{+0.02}$ &   $110_{-24}^{+31}$ \\ 
Cyg~X-1 & P94108 & 31.1  &   $0.94_{-0.13}^{+0.14}$ &   $102_{-22}^{+22}$ &   $0.18_{-0.05}^{+0.05}$ &   $1.32_{-0.03}^{+0.03}$ &   $362_{-160}^{+286}$ \\ 
GX~339-4 & P92035 & 19.2  &   $0.99_{-0.03}^{+0.13}$ &   $33.9_{-2.7}^{+3.5}$ &   $0.45_{-0.07}^{+0.07}$ &   $1.04_{-0.03}^{+0.02}$ &   $242_{-88}^{+31}$ \\ 
GX~339-4 & P92035 & 21.4  &   $1.09_{-0.08}^{+0.12}$ &   $35.4_{-3.4}^{+4.2}$ &   $0.60_{-0.09}^{+0.09}$ &   $0.98_{-0.03}^{+0.03}$ &   $196_{-61}^{+58}$ \\ 
GX~339-4 & P92035 & 23.4  &   $1.13_{-0.10}^{+0.10}$ &   $33.0_{-3.0}^{+3.0}$ &   $0.56_{-0.09}^{+0.09}$ &   $0.90_{-0.03}^{+0.03}$ &   $219_{-59}^{+80}$ \\ 
GX~339-4 & P92052 & 8.95  &   $1.12_{-0.02}^{+0.21}$ &   $53.9_{-7.0}^{+10}$ &   $0.420_{-0.07}^{+0.07}$ &   $1.35_{-0.04}^{+0.03}$ &   $44.7_{-20.0}^{+1.8}$ \\ 
GX~339-4 & P90118 & 6.47  &   $1.33_{-0.07}^{+0.19}$ &   $65.6_{-13}^{+14}$ &   $0.45_{-0.09}^{+0.09}$ &   $1.34_{-0.04}^{+0.04}$ &   $17.8_{-6.6}^{+3.7}$ \\ 
XTE~J1118+480 & P50133 & 3.14  &   $1.01_{-0.15}^{+0.15}$ &   $67.4_{-18}^{+13}$ &   $0.00_{-0.00}^{+0.10}$ &   $1.24_{-0.06}^{+0.04}$ &   $34_{-14}^{+26}$ \\ 
XTE~J1118+480 & P50133 & 3.66  &   $1.02_{-0.10}^{+0.10}$ &   $71.5_{-11}^{+11}$ &   $0.01_{-0.01}^{+0.04}$ &   $1.27_{-0.03}^{+0.03}$ &   $34.8_{-10}^{+15}$ \\ 
XTE~J1118+480 & P50133 & 3.33  &   $0.840_{-0.08}^{+0.16}$ &   $46_{-10}^{+10}$ &   $0.02_{-0.02}^{+0.06}$ &   $1.21_{-0.03}^{+0.03}$ &   $70_{-33}^{+32}$ \\ 
XTE~J1118+480 & P50133 & 3.43  &   $0.910_{-0.15}^{+0.15}$ &   $53.1_{-9.1}^{+9.1}$ &   $0.00_{-0.00}^{+0.04}$ &   $1.22_{-0.03}^{+0.03}$ &   $53_{-23}^{+52}$ \\ 
XTE~J1118+480 & P50133 & 3.33  &   $0.96_{-0.22}^{+0.16}$ &   $50.2_{-9.7}^{+9.7}$ &   $0.01_{-0.01}^{+0.06}$ &   $1.20_{-0.03}^{+0.03}$ &   $41_{-17}^{+70}$ \\ 
XTE~J1118+480 & P50133 & 3.51  &   $1.02_{-0.09}^{+0.09}$ &   $56.3_{-9.9}^{+9.9}$ &   $0.01_{-0.01}^{+0.06}$ &   $1.22_{-0.03}^{+0.03}$ &   $34.7_{-9.8}^{+14.0}$ \\ 
XTE~J1118+480 & P50133 & 3.48  &   $1.02_{-0.11}^{+0.11}$ &   $57.0_{-9.0}^{+9.0}$ &   $0.01_{-0.01}^{+0.05}$ &   $1.23_{-0.03}^{+0.03}$ &   $35_{-12}^{+18}$ \\ 
GRO~J1654-40 & P91702 & 8.55  &   $0.90_{-0.05}^{+0.54}$ &   $83.8_{-19}^{+19}$ &   $0.17_{-0.09}^{+0.09}$ &   $1.48_{-0.08}^{+0.08}$ &   $97_{-78}^{+21}$ \\ 
GRO~J1654-40 & P91702 & 11.3  &   $0.93_{-0.10}^{+0.24}$ &   $122_{-18}^{+18}$ &   $0.19_{-0.08}^{+0.08}$ &   $1.29_{-0.04}^{+0.04}$ &   $146_{-84}^{+84}$ \\ 
GRO~J1654-40 & P91702 & 13.0  &   $0.94_{-0.09}^{+0.12}$ &   $103_{-10}^{+10}$ &   $0.38_{-0.06}^{+0.06}$ &   $1.11_{-0.02}^{+0.02}$ &   $188_{-69}^{+97}$ \\ 
GRO~J1654-40 & P91702 & 13.7  &   $0.87_{-0.12}^{+0.15}$ &   $103_{-15}^{+15}$ &   $0.38_{-0.09}^{+0.09}$ &   $1.05_{-0.03}^{+0.03}$ &   $298_{-138}^{+236}$ \\ 
\hline
4U~1608-52 & P60052 & 2.63  &   $0.93_{-0.09}^{+0.09}$ &   $61.6_{-12}^{+12}$ &   $0.51_{-0.16}^{+0.16}$ &   $0.77_{-0.03}^{+0.03}$ &   $68_{-20}^{+29}$ \\ 
4U~1608-52 & P60052 & 2.27  &   $0.84_{-0.07}^{+0.07}$ &   $77.2_{-9.4}^{+9.4}$ &   $0.37_{-0.17}^{+0.17}$ &   $0.78_{-0.03}^{+0.03}$ &   $91_{-27}^{+38}$ \\ 
4U~1608-52 & P60052 & 2.31  &   $0.93_{-0.06}^{+0.06}$ &   $82.4_{-12}^{+12}$ &   $0.47_{-0.15}^{+0.15}$ &   $0.77_{-0.03}^{+0.03}$ &   $61_{-14}^{+18}$ \\ 
4U~1608-52 & P60052 & 2.36  &   $0.87_{-0.07}^{+0.07}$ &   $80.2_{-15}^{+15}$ &   $0.47_{-0.18}^{+0.18}$ &   $0.76_{-0.03}^{+0.03}$ &   $87_{-23}^{+32}$ \\ 
4U~1608-52 & P60052 & 2.78  &   $0.87_{-0.12}^{+0.13}$ &   $57.6_{-14}^{+16}$ &   $0.45_{-0.15}^{+0.15}$ &   $0.81_{-0.03}^{+0.03}$ &   $85_{-34}^{+57}$ \\ 
4U~1636-536 & P92023 & 2.49  &   $0.97_{-0.01}^{+0.53}$ &   $22.8_{-3.4}^{+26}$ &   $0.18_{-0.14}^{+0.57}$ &   $0.96_{-0.19}^{+0.03}$ &   $44.0_{-34.0}^{+1.0}$ \\ 
4U~1636-536 & P92023 & 2.83  &   $1.40_{-0.20}^{+0.20}$ &   $23.4_{-9.1}^{+11}$ &   $0.45_{-0.37}^{+0.37}$ &   $0.81_{-0.09}^{+0.09}$ &   $13.7_{-5.1}^{+8.2}$ \\ 
4U~1636-536 & P92023 & 3.13  &   $1.50_{-0.19}^{+0.19}$ &   $25_{-10}^{+10}$ &   $0.80_{-0.39}^{+0.39}$ &   $0.76_{-0.09}^{+0.09}$ &   $11.6_{-4.1}^{+6.3}$ \\ 
4U~1636-536 & P92023 & 3.65  &   $1.59_{-0.25}^{+0.25}$ &   $32_{-14}^{+14}$ &   $0.57_{-0.31}^{+0.31}$ &   $0.80_{-0.09}^{+0.09}$ &   $10.7_{-4.5}^{+7.6}$ \\ 
4U~1636-536 & P92023 & 3.56  &   $1.32_{-0.11}^{+0.51}$ &   $15.1_{-2.4}^{+14.0}$ &   $0.18_{-0.18}^{+0.61}$ &   $0.94_{-0.24}^{+0.06}$ &   $21.1_{-14.0}^{+8.1}$ \\ 
4U~1636-536 & P94310 & 2.17  &   $1.23_{-0.25}^{+0.25}$ &   $18.1_{-6.7}^{+9.2}$ &   $0.38_{-0.37}^{+0.41}$ &   $0.75_{-0.11}^{+0.11}$ &   $20.5_{-10}^{+20}$ \\ 
4U~1705-44 & P20073 & 2.73  &   $1.98_{-0.10}^{+0.02}$ &   $26.6_{-5.5}^{+5.5}$ &   $0.75_{-0.35}^{+0.35}$ &   $0.62_{-0.07}^{+0.08}$ &   $4.23_{-0.27}^{+0.76}$ \\ 
4U~1705-44 & P20073 & 2.52  &   $1.97_{-0.14}^{+0.03}$ &   $20.9_{-5.9}^{+5.7}$ &   $1.01_{-0.50}^{+0.49}$ &   $0.56_{-0.10}^{+0.10}$ &   $4.40_{-0.33}^{+1.2}$ \\ 
4U~1728-33& P92023 & 6.69  &   $2.00_{-0.07}^{+0.00}$ &   $18.9_{-2.9}^{+2.9}$ &   $0.52_{-0.25}^{+0.25}$ &   $0.72_{-0.07}^{+0.07}$ &   $9.03_{-0.53}^{+0.98}$ \\ 
4U~1728-33& P92023 & 6.97  &   $1.99_{-0.05}^{+0.01}$ &   $12.8_{-1.2}^{+1.2}$ &   $0.29_{-0.25}^{+0.25}$ &   $0.81_{-0.08}^{+0.08}$ &   $9.48_{-0.39}^{+0.81}$ \\ 
Aql~X-1 & P50049 & 5.85  &   $1.56_{-0.14}^{+0.16}$ &   $26.2_{-6.8}^{+7.7}$ &   $0.57_{-0.26}^{+0.27}$ &   $0.80_{-0.07}^{+0.07}$ &   $18.4_{-5.1}^{+7.0}$ \\ 
Aql~X-1 & P91414 & 2.37  &   $1.09_{-0.15}^{+0.24}$ &   $30.4_{-11}^{+15}$ &   $0.69_{-0.34}^{+0.38}$ &   $0.82_{-0.08}^{+0.08}$ &   $28_{-14}^{+20}$ \\ 
Aql~X-1 & P91414 & 3.68  &   $1.46_{-0.16}^{+0.16}$ &   $25.4_{-6.8}^{+6.8}$ &   $1.11_{-0.36}^{+0.36}$ &   $0.66_{-0.07}^{+0.07}$ &   $17.6_{-5.6}^{+8.3}$ \\ 
Aql~X-1 & P91414 & 3.51  &   $1.40_{-0.21}^{+0.21}$ &   $46.5_{-14.0}^{+14.0}$ &   $0.77_{-0.27}^{+0.27}$ &   $0.76_{-0.06}^{+0.06}$ &   $17.7_{-7.2}^{+12.0}$ \\
\end{tabular}
\caption{Results from spectral analysis of BH (top) and NS (bottom)  X-ray binaries for the \compps  \ parameters; best fit location together with 1D $90\%$ confidence.  The $F_{2-200}$ is the unabsorbed $2-200$~keV flux in the units of $10^{-9}~{\rm erg~cm^{-2}~s^{-1}}$; temperatures are given in keV.}
  \label{tab:compps}
\end{table*}

\begin{table*}
  \begin{tabular}{lccccc}
Source & obsID &  $E_{peak}$ & $\sigma$~ & $EW$~ & Norm.   \\ 
\hline 
GS~1354-64 & P20431 &    $6.3_{-0.2}^{+0.2}$ &   $0.36_{-0.29}^{+0.29}$ &   $0.06_{-0.04}^{+0.08}$ &   $0.70_{-0.19}^{+0.25}$ \\ 
GS~1354-64 & P20431 &    $6.4_{-0.5}^{+0.5}$ &   $0.40_{-0.39}^{+0.77}$ &   $0.06_{-0.04}^{+0.07}$ &   $0.60_{-0.10}^{+0.31}$ \\ 
GS~1354-64 & P20431 &    $6.2_{-0.3}^{+0.3}$ &   $0.51_{-0.50}^{+0.50}$ &   $0.06_{-0.04}^{+0.08}$ &   $0.60_{-0.10}^{+0.25}$ \\ 
GS~1354-64 & P20431 &    $6.1_{-0.2}^{+0.2}$ &   $0.64_{-0.42}^{+0.42}$ &   $0.06_{-0.04}^{+0.08}$ &   $0.80_{-0.28}^{+0.42}$ \\ 
4U~1543-47 & P70124 &    $6.2_{-0.3}^{+0.3}$ &   $0.62_{-0.40}^{+0.40}$ &   $0.30_{-0.15}^{+0.46}$ &   $1.40_{-0.64}^{+1.10}$ \\ 
XTE~J1550-564 & P30188 &    $6.6_{-0.2}^{+0.2}$ &   $0.81_{-0.22}^{+0.22}$ &   $0.10_{-0.08}^{+0.13}$ &   $6.4_{-1.3}^{+1.6}$ \\ 
XTE~J1550-564 & P30188 &    $6.6_{-0.2}^{+0.2}$ &   $0.85_{-0.20}^{+0.20}$ &   $0.10_{-0.08}^{+0.12}$ &   $7.5_{-1.3}^{+1.6}$ \\ 
XTE~J1550-564 & P30188 &    $6.7_{-0.1}^{+0.1}$ &   $1.1_{-0.07}^{+0.17}$ &   $0.13_{-0.11}^{+0.15}$ &   $13._{-1.4}^{+1.7}$ \\ 
XTE~J1550-564 & P50134 &    $6.2_{-0.4}^{+0.4}$ &   $1.5_{-0.31}^{+0.31}$ &   $0.37_{-0.29}^{+0.45}$ &   $26._{-7.3}^{+10.}$ \\ 
XTE~J1550-564 & P50135 &    $5.5_{-0.01}^{+0.7}$ &   $2.0_{-0.33}^{+0.02}$ &   $0.52_{-0.43}^{+0.61}$ &   $22_{-9.1}^{+2.5}$ \\ 
XTE~J1550-564 & P50135 &    $5.5_{-0.01}^{+0.8}$ &   $1.9_{-0.6}^{+0.2}$ &   $0.36_{-0.22}^{+0.51}$ &   $11_{-6.4}^{+1.8}$ \\ 
XTE~J1550-564 & P50135 &    $5.7_{-0.2}^{+0.8}$ &   $1.9_{-0.7}^{+0.09}$ &   $0.30_{-0.15}^{+0.46}$ &   $7.7_{-4.9}^{+3.9}$ \\ 
XTE~J1550-564 & P50137 &    $6.5_{-0.2}^{+0.2}$ &   $1.1_{-0.3}^{+0.3}$ &   $0.15_{-0.12}^{+0.19}$ &   $11_{-2.9}^{+3.9}$ \\ 
Cyg~X-1 & P20173 &    $6.3_{-0.1}^{+0.1}$ &   $0.60_{-0.20}^{+0.20}$ &   $0.10_{-0.08}^{+0.12}$ &   $7.3_{-1.6}^{+1.9}$ \\ 
Cyg~X-1 & P20173 &    $6.3_{-0.1}^{+0.2}$ &   $0.61_{-0.23}^{+0.19}$ &   $0.09_{-0.07}^{+0.11}$ &   $8.1_{-2.0}^{+1.6}$ \\ 
Cyg~X-1 & P20173 &    $6.4_{-0.1}^{+0.1}$ &   $0.51_{-0.18}^{+0.18}$ &   $0.06_{-0.05}^{+0.07}$ &   $7.2_{-1.6}^{+1.3}$ \\ 
Cyg~X-1 & P20173 &    $6.4_{-0.1}^{+0.1}$ &   $0.56_{-0.23}^{+0.23}$ &   $0.09_{-0.07}^{+0.11}$ &   $7.6_{-2.4}^{+1.6}$ \\ 
Cyg~X-1 & P20173 &    $6.3_{-0.1}^{+0.1}$ &   $0.56_{-0.17}^{+0.18}$ &   $0.05_{-0.05}^{+0.06}$ &   $8.4_{-1.4}^{+1.7}$ \\ 
Cyg~X-1 & P20173 &    $6.3_{-0.1}^{+0.1}$ &   $0.57_{-0.16}^{+0.15}$ &   $0.09_{-0.08}^{+0.10}$ &   $8.1_{-1.4}^{+1.4}$ \\ 
Cyg~X-1 & P20173 &    $6.4_{-0.1}^{+0.1}$ &   $0.53_{-0.15}^{+0.15}$ &   $0.09_{-0.07}^{+0.10}$ &   $7.0_{-1.4}^{+1.4}$ \\ 
Cyg~X-1 & P80110 &    $6.3_{-0.2}^{+0.2}$ &   $0.70_{-0.22}^{+0.23}$ &   $0.13_{-0.09}^{+0.16}$ &   $15._{-4.6}^{+6.5}$ \\ 
Cyg~X-1 & P80110 &    $6.3_{-0.2}^{+0.2}$ &   $0.57_{-0.28}^{+0.28}$ &   $0.08_{-0.05}^{+0.11}$ &   $7.0_{-2.5}^{+3.8}$ \\ 
Cyg~X-1 & P80110 &    $6.3_{-0.1}^{+0.1}$ &   $0.32_{-0.26}^{+0.26}$ &   $0.07_{-0.05}^{+0.09}$ &   $5.3_{-1.5}^{+2.1}$ \\ 
Cyg~X-1 & P94108 &    $6.3_{-0.3}^{+0.3}$ &   $0.76_{-0.33}^{+0.33}$ &   $0.19_{-0.12}^{+0.26}$ &   $13._{-5.3}^{+8.8}$ \\ 
GX~339-4 & P92035 &    $6.5_{-0.1}^{+0.1}$ &   $0.87_{-0.15}^{+0.15}$ &   $0.17_{-0.15}^{+0.20}$ &   $14._{-2.4}^{+2.7}$ \\ 
GX~339-4 & P92035 &    $6.5_{-0.2}^{+0.2}$ &   $1.1_{-0.2}^{+0.2}$ &   $0.21_{-0.18}^{+0.25}$ &   $18._{-3.9}^{+4.9}$ \\ 
GX~339-4 & P92035 &    $6.6_{-0.1}^{+0.1}$ &   $0.96_{-0.16}^{+0.16}$ &   $0.21_{-0.18}^{+0.24}$ &   $21._{-3.9}^{+4.7}$ \\ 
GX~339-4 & P92052 &    $6.4_{-0.2}^{+0.2}$ &   $0.68_{-0.38}^{+0.37}$ &   $0.09_{-0.07}^{+0.12}$ &   $2.7_{-1.3}^{+0.88}$ \\ 
GX~339-4 & P90118 &    $6.5_{-0.3}^{+0.3}$ &   $0.050_{-0.040}^{+0.73}$ &   $0.19_{-0.16}^{+0.22}$ &   $0.97_{-0.41}^{+0.73}$ \\ 
XTE~J1118+480 & P50133 &    $6.4_{-0.4}^{+0.4}$ &   $0.75_{-0.51}^{+0.50}$ &   $0.08_{-0.05}^{+0.11}$ &   $0.98_{-0.46}^{+0.86}$ \\ 
XTE~J1118+480 & P50133 &    $6.3_{-0.3}^{+0.3}$ &   $0.75_{-0.41}^{+0.41}$ &   $0.11_{-0.06}^{+0.16}$ &   $1.1_{-0.42}^{+0.71}$ \\ 
XTE~J1118+480 & P50133 &    $6.3_{-0.3}^{+0.3}$ &   $0.65_{-0.49}^{+0.49}$ &   $0.07_{-0.04}^{+0.10}$ &   $0.74_{-0.31}^{+0.54}$ \\ 
XTE~J1118+480 & P50133 &    $6.2_{-0.3}^{+0.3}$ &   $0.30_{-0.29}^{+0.55}$ &   $0.05_{-0.03}^{+0.07}$ &   $0.56_{-0.24}^{+0.41}$ \\ 
XTE~J1118+480 & P50133 &    $6.4_{-0.6}^{+0.6}$ &   $0.24_{-0.23}^{+1.10}$ &   $0.07_{-0.03}^{+0.11}$ &   $0.39_{-0.090}^{+0.84}$ \\ 
XTE~J1118+480 & P50133 &    $6.5_{-0.4}^{+0.4}$ &   $0.55_{-0.53}^{+0.63}$ &   $0.08_{-0.04}^{+0.12}$ &   $0.53_{-0.23}^{+0.46}$ \\ 
XTE~J1118+480 & P50133 &    $6.3_{-0.4}^{+0.4}$ &   $0.61_{-0.59}^{+0.59}$ &   $0.07_{-0.03}^{+0.10}$ &   $0.61_{-0.29}^{+0.55}$ \\ 
GRO~J1654-40 & P91702 &    $5.6_{-0.1}^{+0.9}$ &   $1.5_{-0.7}^{+0.5}$ &   $0.21_{-0.06}^{+0.36}$ &   $7.9_{-6.2}^{+3.7}$ \\ 
GRO~J1654-40 & P91702 &    $5.7_{-0.2}^{+0.9}$ &   $1.7_{-0.}^{+0.3}$ &   $0.35_{-0.18}^{+0.52}$ &   $15_{-8.9}^{+6.9}$ \\ 
GRO~J1654-40 & P91702 &    $5.7_{-0.2}^{+0.5}$ &   $2.0_{-0.2}^{+0.0}$ &   $0.54_{-0.45}^{+0.64}$ &   $26_{-7.1}^{+4.9}$ \\ 
GRO~J1654-40 & P91702 &    $5.6_{-0.1}^{+0.6}$ &   $1.9_{-0.3}^{+0.1}$ &   $0.49_{-0.36}^{+0.61}$ &   $27_{-10}^{+7.7}$ \\ 
\hline
4U~1608-52 & P60052 &    $6.5_{-0.7}^{+0.5}$ &   $1.8_{-0.5}^{+0.3}$ &   $0.12_{-0.05}^{+0.18}$ &   $1.3_{-0.70}^{+1.5}$ \\ 
4U~1608-52 & P60052 &    $5.5_{-0.01}^{+1.10}$ &   $0.0_{-0.0}^{+1.9}$ &   $0.05_{-0.03}^{+0.07}$ &   $0.34_{-0.04}^{+0.22}$ \\ 
4U~1608-52 & P60052 &    $7.0_{-1.3}^{+0.09}$ &   $1.6_{-0.4}^{+0.4}$ &   $0.07_{-0.03}^{+0.11}$ &   $0.42_{-0.12}^{+0.46}$ \\ 
4U~1608-52 & P60052 &    $6.4_{-0.4}^{+0.4}$ &   $0.85_{-0.42}^{+0.42}$ &   $0.12_{-0.06}^{+0.17}$ &   $1.1_{-0.5}^{+0.9}$ \\ 
4U~1608-52 & P60052 &    $5.6_{-0.1}^{+1.2}$ &   $0.07_{-0.07}^{+1.8}$ &   $0.06_{-0.02}^{+0.10}$ &   $0.34_{-0.04}^{+0.72}$ \\ 
4U~1636-536 & P92023 &    $5.8_{-0.3}^{+1.1}$ &   $1.6_{-0.58}^{+0.38}$ &   $0.23_{-0.08}^{+0.39}$ &   $4.5_{-3.5}^{+2.4}$ \\ 
4U~1636-536 & P92023 &    $6.8_{-0.5}^{+0.3}$ &   $0.29_{-0.29}^{+1.3}$ &   $0.09_{-0.01}^{+0.16}$ &   $0.74_{-0.52}^{+1.8}$ \\ 
4U~1636-536 & P92023 &    $6.8_{-0.3}^{+0.3}$ &   $0.16_{-0.16}^{+0.80}$ &   $0.08_{-0.03}^{+0.13}$ &   $0.83_{-0.46}^{+1.00}$ \\ 
4U~1636-536 & P92023 &    $6.8_{-0.4}^{+0.3}$ &   $0.57_{-0.57}^{+1.1}$ &   $0.15_{-0.03}^{+0.27}$ &   $1.3_{-0.86}^{+2.7}$ \\ 
4U~1636-536 & P92023 &    $6.5_{-0.7}^{+0.6}$ &   $1.5_{-1.1}^{+0.5}$ &   $0.08_{-0.01}^{+0.15}$ &   $3.5_{-3.4}^{+2.8}$ \\ 
4U~1636-536 & P94310 &    $6.7_{-0.5}^{+0.4}$ &   $1.1_{-0.6}^{+0.6}$ &   $0.13_{-0.06}^{+0.21}$ &   $2.2_{-1.4}^{+3.6}$ \\ 
4U~1705-44 & P20073 &    $6.1_{-0.20}^{+0.20}$ &   $1.0_{-0.2}^{+0.2}$ &   $0.23_{-0.18}^{+0.29}$ &   $3.2_{-0.9}^{+1.2}$ \\ 
4U~1705-44 & P20073 &    $6.1_{-0.25}^{+0.25}$ &   $1.1_{-0.3}^{+0.3}$ &   $0.23_{-0.17}^{+0.28}$ &   $3.4_{-0.9}^{+1.3}$ \\ 
4U~1728-33& P92023 &    $5.9_{-0.4}^{+0.4}$ &   $1.2_{-0.3}^{+0.3}$ &   $0.31_{-0.22}^{+0.40}$ &   $10._{-3.4}^{+5.1}$ \\ 
4U~1728-33& P92023 &    $5.5_{-0.3}^{+0.3}$ &   $1.4_{-0.2}^{+0.2}$ &   $0.40_{-0.29}^{+0.52}$ &   $19._{-5.5}^{+7.7}$ \\ 
Aql~X-1 & P50049 &    $6.7_{-0.2}^{+0.2}$ &   $0.66_{-0.25}^{+0.25}$ &   $0.09_{-0.05}^{+0.12}$ &   $2.5_{-1.0}^{+1.5}$ \\ 
Aql~X-1 & P91414 &    $6.1_{-0.6}^{+0.7}$ &   $1.2_{-0.7}^{+0.7}$ &   $0.08_{-0.03}^{+0.12}$ &   $1.3_{-0.9}^{+1.8}$ \\ 
Aql~X-1 & P91414 &    $7.0_{-0.5}^{+0.1}$ &   $0.79_{-0.52}^{+0.53}$ &   $0.10_{-0.05}^{+0.15}$ &   $1.4_{-0.8}^{+1.6}$ \\ 
Aql~X-1 & P91414 &    $6.8_{-0.9}^{+0.3}$ &   $1.4_{-0.6}^{+0.6}$ &   $0.15_{-0.04}^{+0.25}$ &   $2.0_{-1.3}^{+4.0}$ \\

\end{tabular}
  \caption{Results from spectral analysis of BH (top) and NS (bottom) X-ray binaries   for the Gaussian component; best fit location together with 1D $90\%$ confidence. The $E_{peak}$, $\sigma$ and $EW$ are in keV, and the normalisation (line flux) in $10^{-3}~{\rm phot~cm^{-2}~s^{-1}}$.}
    \label{tab:line}
\end{table*}

\begin{figure*}
\begin{center}
\includegraphics[width=0.48\textwidth]{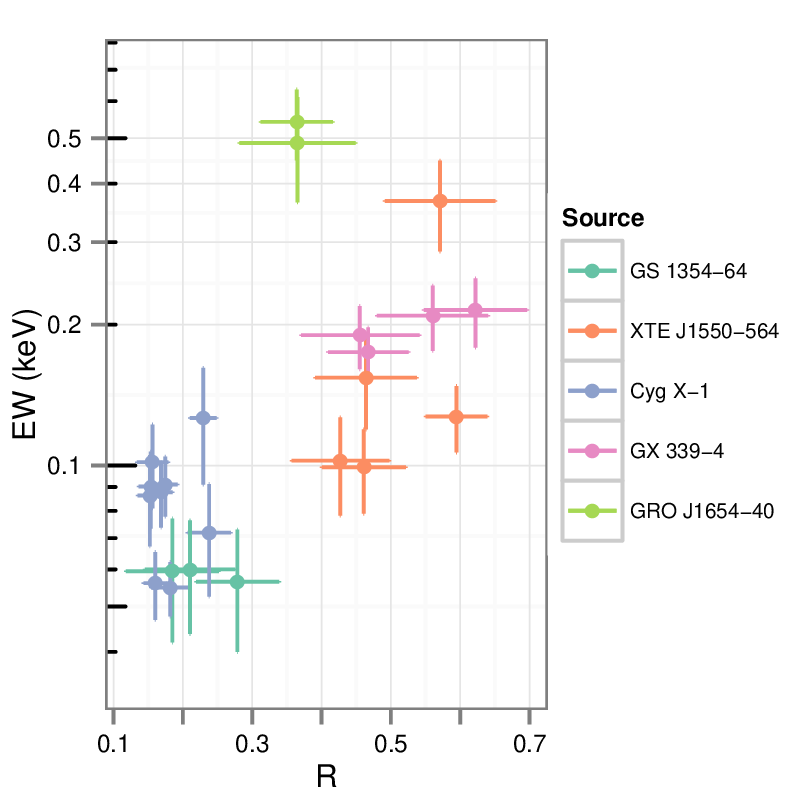}  \includegraphics[width=0.48\textwidth]{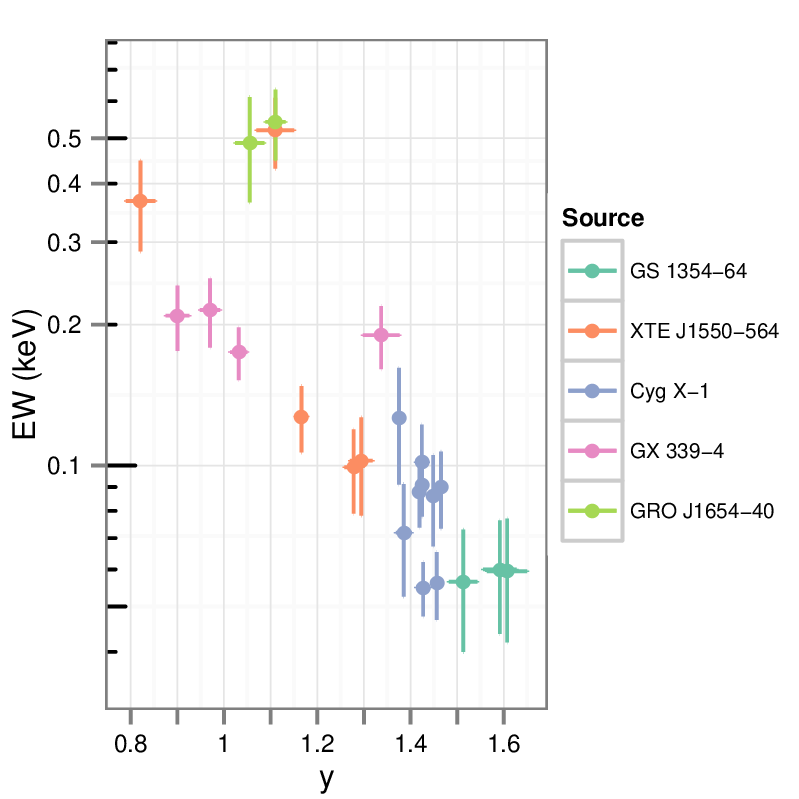}
\caption{The variation of Gaussian equivalent width with reflection strength (left) and Compton $y-$parameter (right) for black hole sources.  We exclude points where \change{$\sigma_{EW}/EW>0.3$}, and also points where $\sigma_R / R > 0.3$ or $\sigma_y / y > 0.3$, respectively\label{fig:BHline}.}
\end{center}
\end{figure*}

\section{Discussion}

\section{The $R-y$ plane}

There is a strong degeneracy between $y$ and $R$, which is similar to the well-known degeneracy between $R$ and photon index $\Gamma$ \citep{1999MNRAS.303L..11Z,1999A&A...352..182G}. This degeneracy coupled with the relatively large confidence intervals means that no obvious trend can be seen for NS systems on the $y-R$ plane.
Most spectra for a particular NS occupy a similar position in the parameter space, showing no clear evolution.  This is to be expected because these sources spend most of their time in the soft (banana) state, which transitions to a low/hard state where the data are of lower quality, leading to larger uncertainties in spectral fitting. The transition to the high-intensity soft state occurs on such short timescales that it is unlikely to observe an Atoll source over a range of intensities while the spectra are hard.  In contrast there is some notable evolution in the $R-y$ plane for several of the BH sources, most notably 4U~$1550-56$, GX~$339-4$ and XTE~J$1654-40$.  When the spectra are taken from the same outburst where the source rises in intensity through the hard state, near to the point where the source transitions; this leads to an increased $R$ and a reduction in $y$.  This behaviour is of the same nature as the $R-\Gamma$ correlations in X-ray binaries and AGN observed previously \citep{1999MNRAS.303L..11Z, 1999A&A...352..182G, 2000sgwa.work..114G,2003MNRAS.342..355Z}, albeit with  larger scatter.

In our spectral model there are two independent measures of the strength of reprocessed emission; the reflection strength $R$($=\Omega /2\pi $) in the Comptonising component and the equivalent width of the Gaussian line used to model the Fe $K\alpha$ emission.  While it is likely that a Gaussian is a poor description of the line profile, the moderate spectral resolution of the PCA instrument means that it is adequate for the purpose of characterising the emission strength.  Using similar spectral models, it has been shown that $R$ and the equivalent width $EW$ of the line correctly rank the spectra in terms of the relative amount of reprocessed emission contributing to the spectrum \citep[e.g.,][]{1999A&A...352..182G}. In figure~\ref{fig:BHline} we show that there is a reasonable correlation between the $EW$ and $R$ when the two parameters are well constrained (which includes the majority of BH sources but not NS sources).  This demonstrates that the spectral modelling returns a consistent measure of the relative amount of reprocessed emission present in a spectrum, even if the exact value of $R$ is systematically under- or over-estimated as a result of not accounting for certain physical variables, such as the ionisation state of the reprocessing material. We also find a clear  anti-correlation between $y$ and $EW$. Interestingly, it has much smaller scatter, than relation between $y$ and $R$, and in this respect is more similar to the previously observed $R-\Gamma$ correlations.

\begin{figure*}
\begin{center}
\includegraphics[width=0.495\textwidth,height=0.5\textwidth]{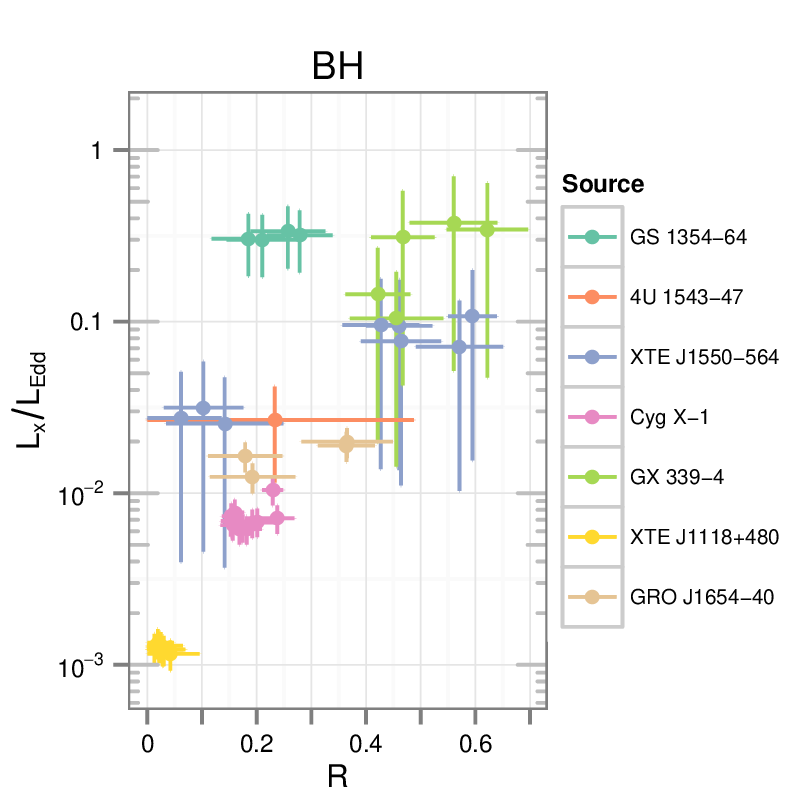}   \includegraphics[width=0.495\textwidth,height=0.5\textwidth]{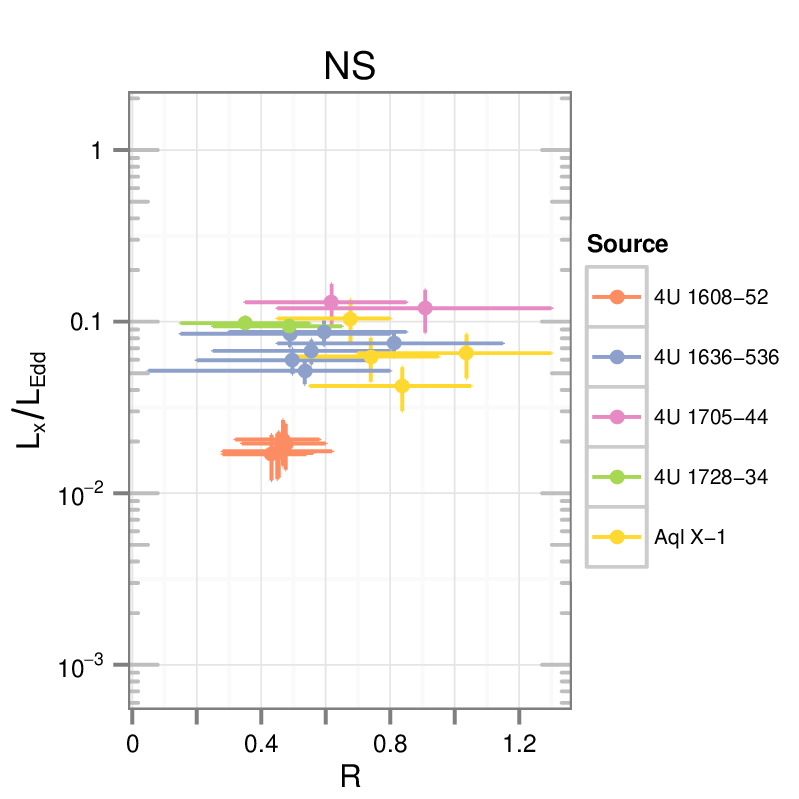}
\caption{The variation of reflection strength with luminosity for BHs (left) and NSs (right).  In the calculation of $L_x/L_{Edd}$ we have assumed the distances and BH masses given in table~\ref{tab:nh}, and a mass of $1.4~\Msol$ for NSs.  We use $15~\Msol$ as a reasonable upper-limit on BH mass when only a lower-limit is available\label{fig:rledd}.}
\end{center}
\end{figure*}

In figure~\ref{fig:rledd} we plot $R$ against luminosity, indicating a trend of increasing $R$ with accretion rate. The highest luminosities are consistent with $\approx (0.1-0.3) L_{Edd}$ for both BH and NS samples (we note that the large uncertainties in $L$ for some sources originate from  the large uncertainty in the distances to some sources).  For NSs the connection between luminosity and reflection is less clear, though this is more difficult to ascertain because of the lack of luminosity variation in the NS hard state compared to that of BH XBs.  The \change{trend between} $R$ with $L/L_{Edd}$ is not mirrored by the Compton $y-$parameter, for which much larger scatter is observed.

\subsection{Comptonisation properties of LMXBs}
\label{subsec:yparm}

\begin{figure*}
\begin{center}
\includegraphics[width=0.49\textwidth]{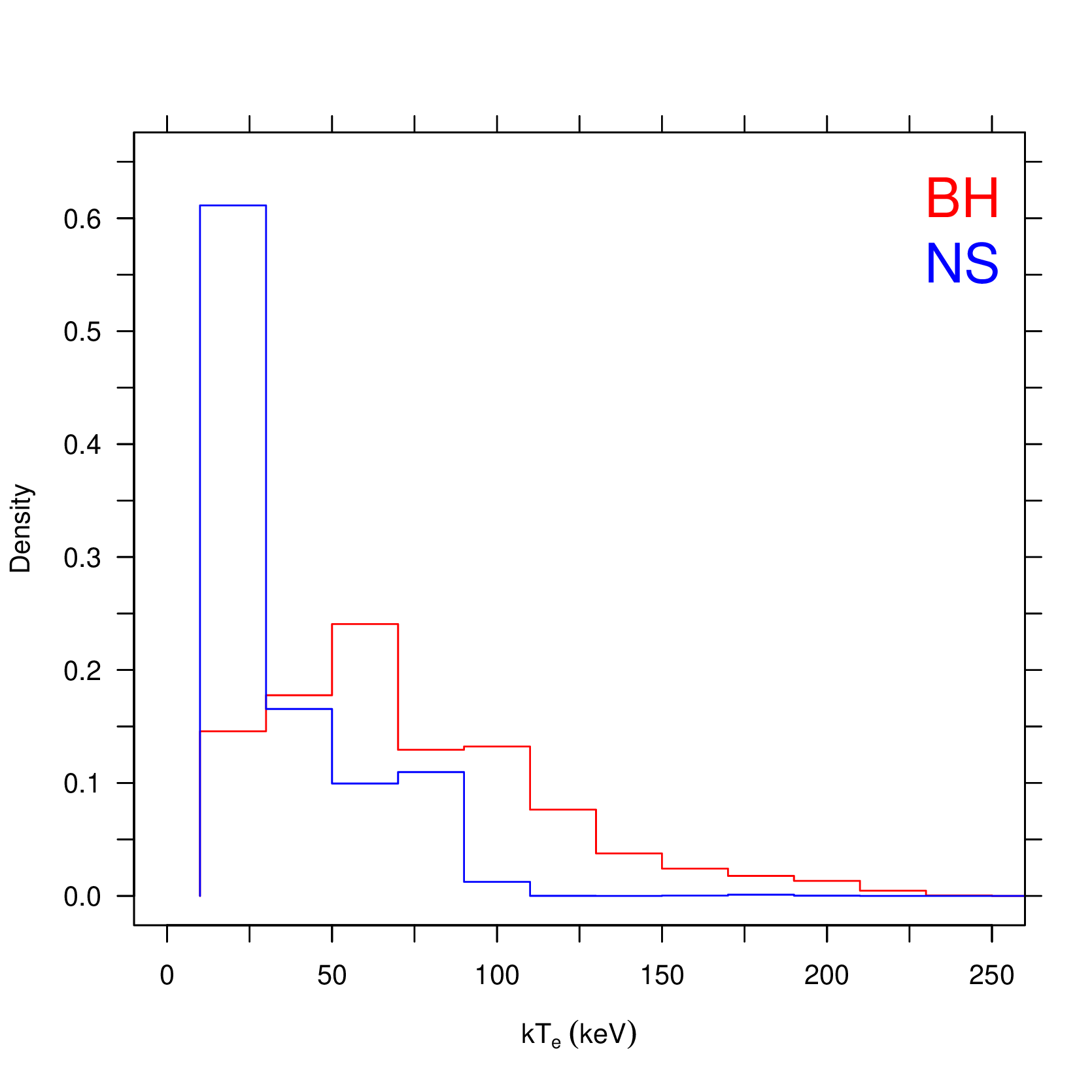} \includegraphics[width=0.49\textwidth]{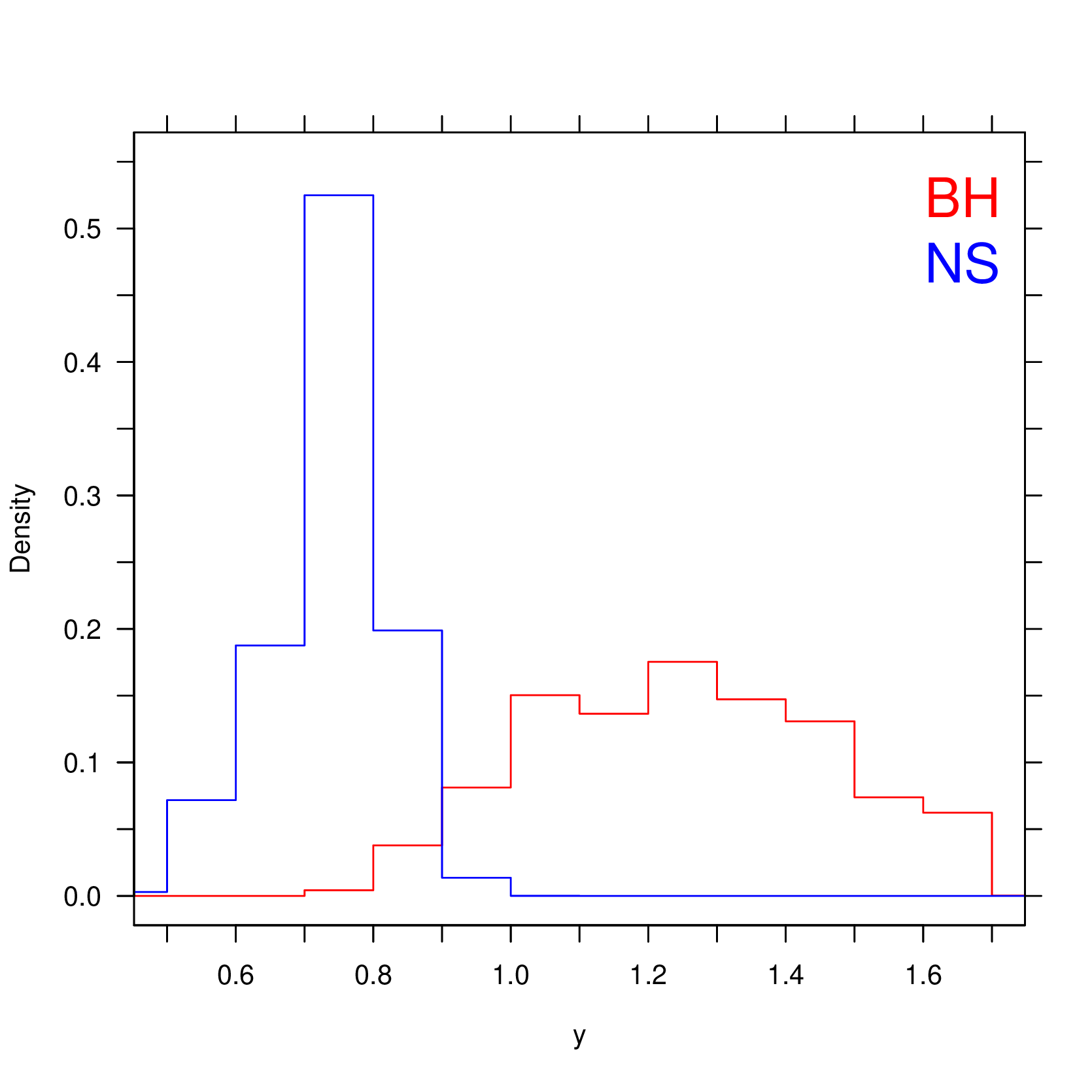}
\caption{The  1D posterior distribution of $kT_e$ (left) and Compton $y-$parameter (right) for BHs and NSs, as defined by equation~\ref{eq:cum}\label{fig:kTedist}.}
\end{center}
\end{figure*}

As one can see from figure \ref{fig:NSBH}, there is a clear dichotomy in $y$ between the NS and BH groups, divided between $y\approx0.9$.  This is excellent qualitative support for the hypothesis that the electron cloud properties in NSs will be affected by additional seed photons from the neutron star, and that this will be observable in the emergent spectrum \citep{1989ESASP.296..627S}.  In figure~\ref{fig:kTedist} we plot the normalised 1D posterior distributions of $kT_e$ and $y$ for NSs and BHs,
\begin{equation}
P(kT_e)=\frac{\sum_{j}^{N_{sources}} \left( \frac{\sum_{i}^{N_{j}} P_{ij}(kT)}{N_{j}} \right)}{N_{sources}},
\label{eq:cum}
\end{equation}
where $N_j$ is the number of spectra for a given source and $P_{ij}(kT)$ denotes the 1D posterior distribution of $kT$ for the $i^{\rm{th}}$ spectrum of source $j$. Figure~\ref{fig:kTedist} highlights the well known dearth of higher temperature ($>50~keV$) electron clouds in NS XBs which $kT_e$ distribution peaks at $\approx 15-25$ keV.  BH systems, on the other hand,  cover a greater range of observed $kT_e$ extending to $kT_e\sim 200$ keV. Interestingly, BH systems can be also observed at similar low $kT_e$ as seen for NSs. The distribution of NS systems over $y$-parameter is also  narrow and strongly peaked at $y\approx 0.7-0.8$, whereas BH systems are  rather broadly distributed  between $y\approx 0.9-1.7$. Figure~\ref{fig:kTedist} illustrates that there is a clear dichotomy in the $y$-parameter distribution with little overlap between BH and NS systems.  In figure~\ref{fig:kTetau} \change{we demonstrate that for a given optical depth the coronae in BH XBs will be at a higher temperature than that measured in NS XBs}.  Alternatively, it can be expressed in terms of BH coronae having larger optical depth at the same temperature.

Several assumptions were made during spectral analysis, including freezing some parameters at a fixed value for all sources.  However, it is unlikely that this greatly affects our conclusions.  The inclination $i$ is not well-constrained for a given source unless \emph{in extremis}, for which dips or eclipses would be observed, but this is also when the inclination would have the greatest effect on our results \citep[some studies suggest a dependence on $y$ with inclination][]{2008PASJ...60..585M,2014PASJ...66..120Z}. \change{ Therefore the lack of extreme inclinations in our sample means that it is unlikely that the observed difference in $y$ is the result of some systematic difference in inclination between the BH and NS groups.}   To see if any dependence exists in our results, it is instructive to compare the analyses of Cyg~X$-1$ and GRO~J1654-40, which are known to have $i\approx45^\degree$ and $i\approx70^\degree$ respectively. We find no firm evidence of a dependence between $y$ and $i$ in the current work.  Two of the GRO~J1654-40 spectra are consistent with those of Cyg X-1 in terms of $y$ (figure~\ref{fig:NSBH}), while two have significantly smaller $y$, and larger $R$.  The two GRO~J1654-40 spectra that are consistent with those of Cyg X-1 in terms of $y$ and $R$ are at a higher $L/L_{Edd}$ (Table \ref{tab:compps}), and this fits with simple geometric reasoning that a higher $y$ (lower $R$) should be observed for a given luminosity the greater the inclination of the source.

\subsection{Compton Amplification}
\label{subsec:compamp}

We calculate a Compton amplification factor $A$, which we define as the ratio of the measured luminosity in the $3-200$ keV band to the seed photon luminosity $L_{seed}$, which we calculate from the seed photon temperature $kT_{bb}$ and \compps~normalisation $N$, for a source at distance $D$

\begin{equation}
L_{seed}=4\pi N D^2 \sigma T_{bb}^4.
\end{equation} 
Note that with this definition, $A$ is only a proxy to the true value of the Compton amplification factor. The main sources of inaccuracy are:   (i) the limited energy range in which the luminosity of the Comptonised component is calculated; (ii) lack of the correction for the interstellar absorption; (iii) a bias in measuring the temperature and normalisation of the seed photons spectrum caused by the  limited low energy coverage of the  RXTE data (see section \ref{subsec:seed}). However, as shown in Section \ref{subsec:seed}, these factors introduce a rather uniform downward bias in $A$,  affecting BH and NS systems in a similar way. The values of  Compton amplification factor computed using the broad band data  (combined XMM-Newton and RXTE) and corrected for the interstellar absorption are consistent with the values reported here within  a factor of $\approx 1.4$.

In figure~\ref{fig:ampy} we plot the Compton $y-$parameter against $A$, following a positive trend, which is expected as both quantities are a direct proxy for the Comptonising ability of the corona.  We find that no NSs have $A > 3$, while BHs occupy a range of $A\sim2.8-7$,  as it is further illustrated in figure~\ref{fig:amp_hist},  where we plot the distribution of sources over the amplification factor, separately for NS and BH systems.

\begin{figure}
\begin{center}
\includegraphics[width=0.48\textwidth]{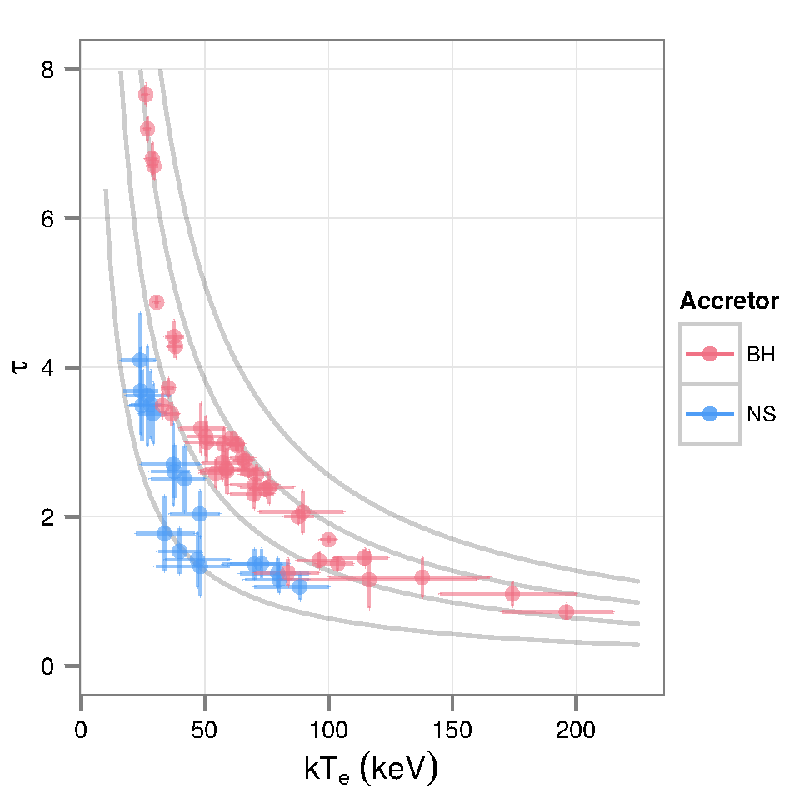}
\caption{Calculated optical depth against electron temperature for all NS and BH spectra, overlaid on lines of constant Compton $y$-parameter, where $y=0.5,1.0,1.5,2.0$ (ascends with $\tau$) \label{fig:kTetau}.}
\end{center}
\end{figure}

To understand the physical implications of our results in terms of \change{the supply of seed photons}, we invoke the following, admittedly very simplistic, considerations. In case of BHs, the Compton amplification factor is (assuming that seed photons are supplied by the accretion disk):
\begin{equation}
A_{BH}=\frac{L_{Compt}}{fL_{disk}}
\label{eq:abh}
\end{equation}
where the factor $f$ accounts for the fact that only some fraction of the disc emission is intercepted by the hot Comptonising corona. The largeness of the $A_{BH}$ ($\sim 4-6$) is qualitatively consistent with the truncated disc picture (note however, that  estimation of the disc truncation radius from eq.(\ref{eq:abh}) requires knowledge of the factor $f$ and the feedback coefficient between the corona and the accretion  disc \citep[see][]{1995ASIC..450..331G}.

In the case of NS, there is an additional potentially powerful source of  soft emission, the surface of the neutron star, therefore the Compton amplification factor equals to:
\begin{equation}
A_{NS}=\frac{L_{Compt}}{f_1L_{disk}+f_2L_{NS}}
\label{eq:ans}
\end{equation}
where, as before, $f_1$ accounts for the fraction of disc photons intercepted by the corona and $f_2$ is a similar fraction for photons emitted by the neutron star. If the NS is surrounded by the (quasi-) spherically symmetric corona, $f_2=1-e^{-\tau_T}\approx 1$, as $\tau_T\ga 1$ (figure \ref{fig:kTetau}).  It is plausible that the decrease of the Compton amplification factor in NS systems is caused by the contribution of the NS luminosity, rather than by  a global change of the accretion geometry, therefore  $f_2L_{NS}\ga f_1L_{disk}$.  For the purpose of this crude estimation we will further ignore  $f_1L_{disk}$ and assume $f_2=1$. Taking into account that $L_{Compt}=W_{corona}+f_2L_{NS}$ and $L_{NS}=W_{NS}$, where $W_{corona}$ and $W_{NS}$ are the energy release in the hot corona and on the NS surface, we can rewrite eq.(\ref{eq:ans}):
\begin{equation}
A_{NS}\approx \frac{W_{corona}+W_{NS}}{W_{NS}} \sim 2-3
\label{eq:ans1}
\end{equation}
from which we obtain:
\begin{equation}
\frac{W_{corona}}{W_{NS}}\sim 1-2
\label{eq:ans2}
\end{equation}
We thus conclude that accreting material looses in the Comptonisation process in the corona about $\sim 1/2-2/3$ of the total energy it possessed upon entering the corona. Remaining energy is released in the form of kinetic energy of the infalling matter  on the surface of the neutron star, thus making it a powerful source of soft seed photons for Comptonisation. This explains systematically lower Compton amplification factor and lower $y$-parameter in the NS systems. It is interesting to note that the fraction of accretion energy released on the NS surface in the hard state is comparable to that in the soft state \citep{1988AdSpR...8..135S}.  It's particular value in the hard state  is controlled by the  efficiency of Comptonisation and coupling between protons and electrons in the hot corona.

\begin{figure}
\begin{center}
\includegraphics[width=0.49\textwidth]{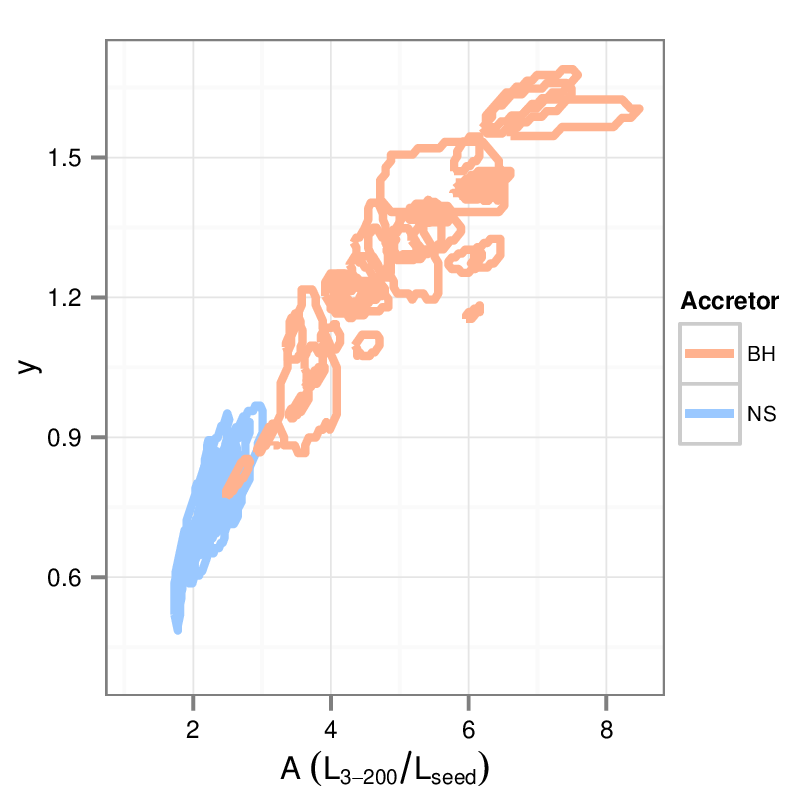}
\caption{Compton y-parameter against Compton amplification for all sources\label{fig:ampy}.}
\end{center}
\end{figure}

\begin{figure}
\begin{center}
\includegraphics[width=0.49\textwidth]{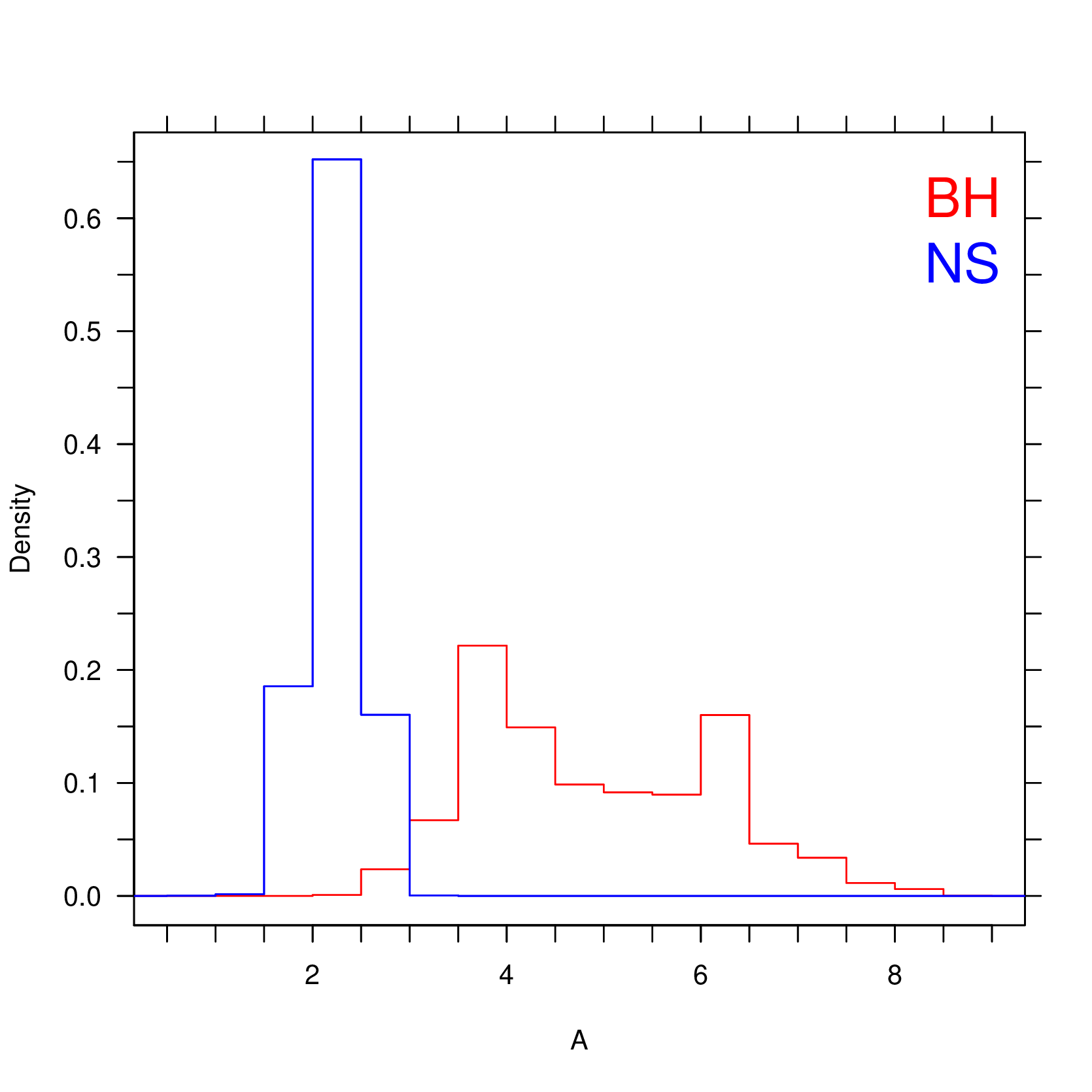}
\caption{\label{fig:fa}The cumulative 1D posterior distribution of $A$, as with figure~\ref{fig:kTedist}.\label{fig:amp_hist}}
\end{center}
\end{figure}

\subsection{Effects of rotation on Comptonisation properties}
\label{subsec:rotation}

\begin{figure*}
\begin{center}
\hbox{
\includegraphics[width=0.43\textwidth]{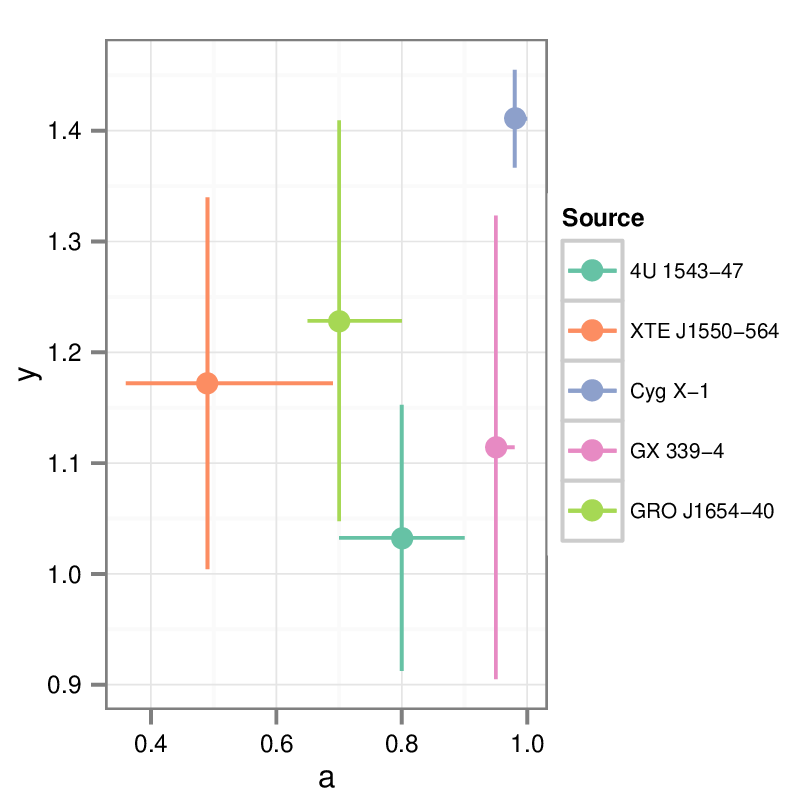} \hspace{1cm}
\includegraphics[width=0.43\textwidth]{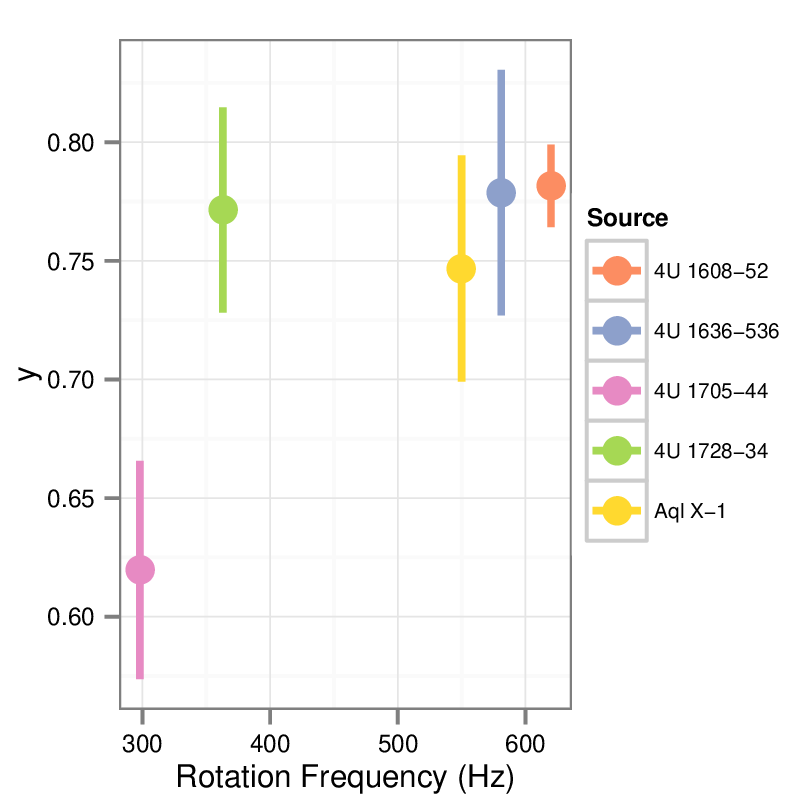} }
\hbox{
\includegraphics[width=0.43\textwidth]{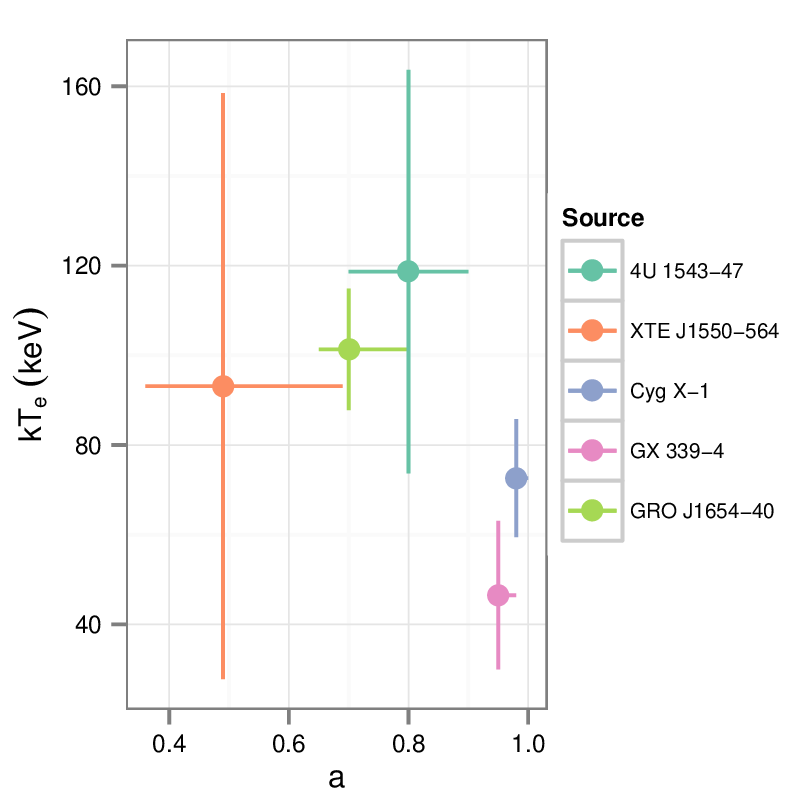} \hspace{1cm}
\includegraphics[width=0.43\textwidth]{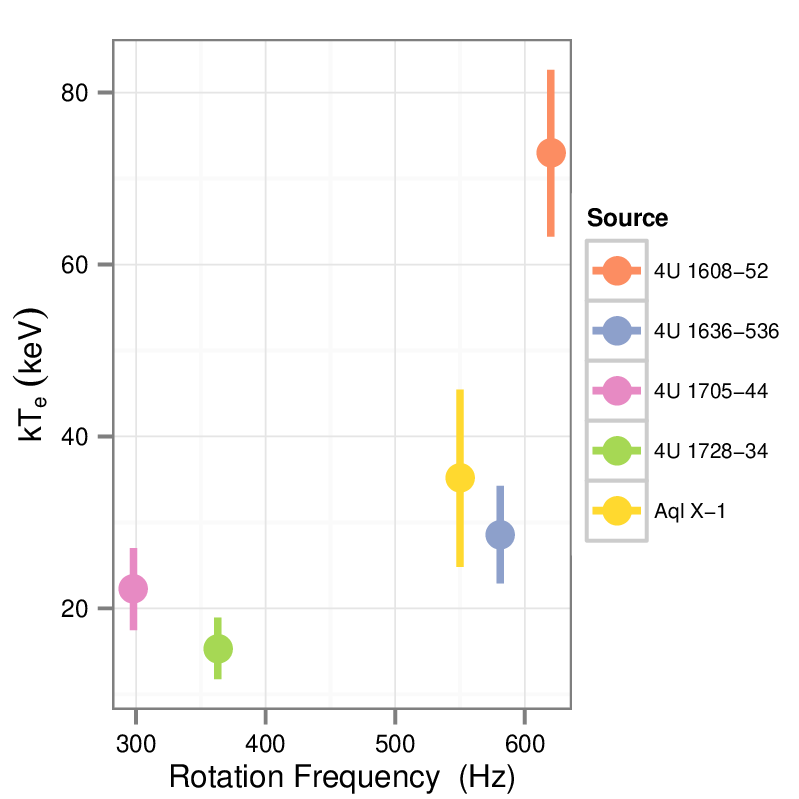}  }
\hbox{
\includegraphics[width=0.43\textwidth]{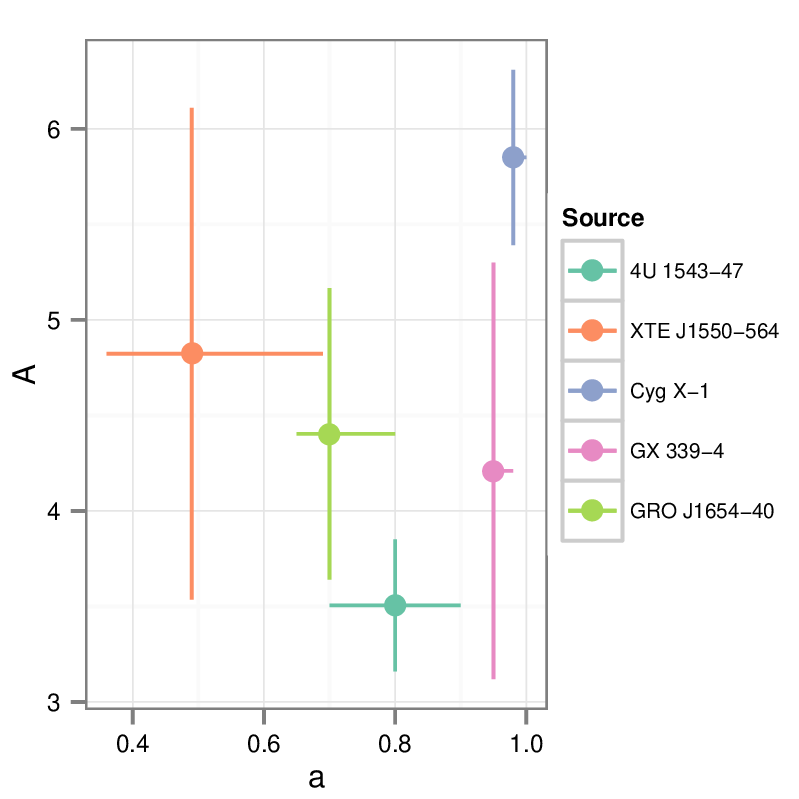} \hspace{1cm}
\includegraphics[width=0.43\textwidth]{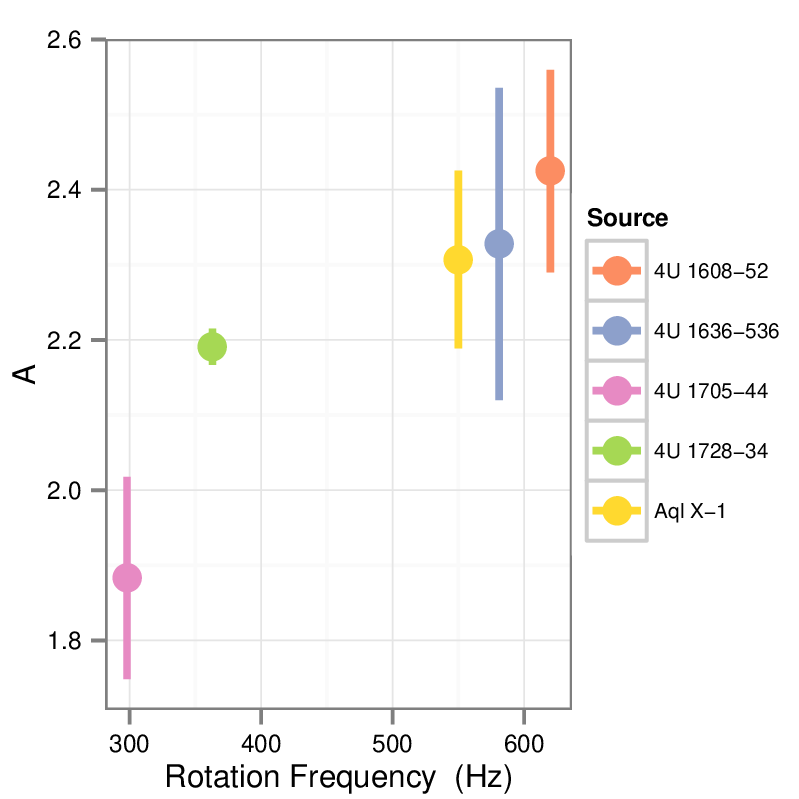}}
\end{center}
\caption{Average of Comptonisation parameters against BH spin (left) and NS rotation frequency (right), error bars represent standard deviation (or $90\%$ confidence in case of only one spectrum).   \label{fig:spins}}
\end{figure*}

It is anticipated that the rotation of a compact object is an important parameter in the context of the spectral formation, and to test this we take the step of plotting the Comptonisation parameters against rotation frequency (for sources for which the rotation is known).  For NS rotation   we use measurements of the burst oscillation frequency \citep[see e.g.][]{1996ApJ...469L...9S} or in one instance the difference in frequency between kHz QPOs \citep[4U~$1705-44$, for which burst oscillations have never been observed,][]{1998ApJ...498L..41F}, while for BHs we use the dimensionless spin parameter, $a$, as found from either studies of the fluorescent iron line profile \citep[see][]{1989MNRAS.238..729F,1991ApJ...376...90L,2003PhR...377..389R}, in the case of $GX~339-4$, or the thermal continuum of the accretion disc \citep{1997ApJ...482L.155Z, 2014SSRv..183..295M} for all other sources.   In figure~\ref{fig:spins} we present the key Comptonisation parameters against compact object rotation.  For each source we defined the mean value for each spectral parameter to be used as a characteristic value.  Black hole spins were obtained from \protect\cite{2015ApJ...813...84G} (GX~$339-4$), \protect\cite{2014grca.book..523N} (GRO~J$1655-40$, 4U~$1543-47$), \protect\cite{2014ApJ...790...29G} (Cyg~$X-1$), \protect\cite{2011MNRAS.416..941S} (XTE~J$1550-564$). Neutron star rotation frequencies were obtained from \protect\cite{2001ApJ...553L.157M} and \protect\cite{1998ApJ...498L..41F} (4U~$1705-44$).  

In the case of NSs, there seems to be some correlation between the spin frequency and the Compton amplification $A$, which may also be present with $y$ and $kT_e$.  The Spearman rank correlation test \citep{10.2307/1412159} gives an estimate of the probability of  appearance of the observed correlations of points by chance of 0.017, 0.08 and 0.13 for $A$, $y$ and $kT_e$ respectively. Therefore, the correlation of A with spin frequency is detected at the confidence level equivalent to $\approx 2.4\sigma$ for a Gaussian distribution. Although this is not enough to claim a confident detection, the probability of such correlation to be real is quite high.

A positive correlation between the strength of Comptonisation and the intrinsic NS rotation frequency is to be expected on theoretical grounds. 
Indeed, for the standard Keplerian accretion disk,  the luminosity of the boundary layer on the NS surface is, in the Newtonian approximation:
\begin{equation}
L_{bl}=\frac{1}{2}\,\dot{M} \left( v_K- v_{NS}\right)^2
\label{eq:lbl}
\end{equation}
where $v_{NS}$ -- the linear velocity of the neutron star surface at the equator and $v_K$ -- Keplerian velocity  near the NS surface \citep{1988AdSpR...8..135S,1988PhDT.........2K,2000AstL...26..699S}. The higher the NS spin (i.e. higher $v_{NS}$)\footnote{note that in all cases where the NS spin is measured, $v_{NS}<v_K$}, the smaller is the $L_{bl}$. As for NS systems $L_{seed}\approx L_{bl}$, and, by definition,  $A=L_{tot}/L_{seed}$, higher spin systems should have smaller $L_{seed}$ and, correspondingly,  larger $A$. In order words, for a slowly rotating  NS, the luminosity of its surface is higher and Comptonisation in the corona is weaker. Copious soft photons are  more efficient in cooling  electrons of the corona, leading to decrease of $kT_e$, $y$ and $A$ \citep{1989ESASP.296..627S}. In the hard state sources considered here the inner accretion flow is hot, geometrically thick and probably, sub-Keplerian, however  qualitatively, the picture outlined above holds.
In addition to being a  confirmation of the theoretical picture, this behaviour also presents the exciting potential of being able to determine a NS spin from the X-ray spectra alone.    

For BHs, there is no such obvious trend between the Comptonisation properties and spin, however, these spins are much less well constrained than NS rotation frequencies and may be subject to systematic uncertainties arising from choice of spectral model.

\subsection{\change{Classifying compact objects from Comptonisation parameters}}
The observed dichotomy \change{between BH and NS LMXBs} in the properties of the Comptonising media has an intriguing potential as a diagnostic.  Conclusive classification of a compact object as a BH requires dynamical mass measurements obtained by optical or IR spectroscopy of transient sources during quiescence, when the optical emission is dominated by the companion \citep[see][for review]{2006csxs.book..215C}.   The necessity for a source to be transient, and for the secondary to be bright enough to study with optical instrumentation, limits our ability to firmly identify BHs.  However, there is an increasing list of behaviours that can be used to identify a system as containing a black hole `candidate' (BHC).  Examples of temporal behaviour displayed by NS LMXBs but not BH LMXBs include type-I X-ray bursts and a \newchange{significant amount of noise in their X-ray variability at frequencies greater than $\sim500$~Hz while in the hard state \citep{2000A&A...358..617S}}.  We can also look to the spectral properties,  as BH sources should be softer during the so-called high state, where the emission from the disc dominates.  Such emission is characterised by the temperature of innermost portion of the disc, the position of which is determined by the ISCO of the compact object, which we naturally expect to be larger for BHs than for NSs ( for which the inner disc radius my be determined by the  NS radius).  The temperature of the inner accretion disc has been used for the classification of BHs in Extragalactic studies \citep[e.g.][]{2010ApJ...725.1805B,2013ApJ...766...88B,2015arXiv150605448L}, where dynamical determination of the mass is nearly impossible and photon-starved X-ray lightcurves limit our ability to identify X-ray bursts but longer observation times yield an adequate spectral quality for analysis. 

From the observed distribution of the \change{the Compton y-parameter} and amplification factor during the  hard state we propose a further characteristic property for distinguishing BHs from NSs.  The population is clearly bimodal about $y\approx0.9$ and  $A\approx3$ (as measured over the $3-200$ keV energy band, see further discussion in \S\ref{subsec:seed}), with some slight overlap (figure~\ref{fig:kTedist}).    Determination of the Compton $y-$parameter can be immediately incorporated into the analysis of newly discovered Galactic XBs and may prove effective for studying populations of Extragalactic XBs using future X-ray telescopes.

\subsubsection{Comparison with other Comptonisation models}
It is prudent to examine whether the observed parameter degeneracy is apparent when using different or more sophisticated Comptonisation models.  For this task we choose {\sc eqpair} \citep[an extension of the code presented in][]{1992MNRAS.258..657C,1999ASPC..161..375C}, which considers a hot plasma within which a hybrid of thermal and non-thermal emission processes take place, while the reflection component is considered by the same routines as \compps.  The model is parametrised in terms of  the ratio of the hard to soft compactnesses $l_h/l_s$, the soft photon compactness $l_s$, seed photon blackbody temperature $kT_{bb}$, Thomson scattering depth $\tau_p$, the fraction of power supplied to energetic particles that goes into accelerating  non-thermal particles $l_{nt}/l_h$  and the reflection fraction $R$.  An aspect of this particular model is that the electron temperature $kT_e$ is not a parameter, but it is calculated self-consistently and can be extracted through setting the \xspec~ chatter level to 15.   

Taking sources whose spectra were typical of those possessing either a BH or NS -- 4U~$1636-536$ from the NSs and Cyg X-1 from the BHs -- and we analysed several spectra for each source  (those from P92023 for 4U~$1636-536$, and from P20173 for Cyg~X-1), but with {\sc eqpair} taking the place of \compps, achieving acceptable fits.  The range of reported $kT_e$ for each source was $15-22$ keV for the NS, and $55-67$ keV for the BH, consistent with our \compps~ results (table~\ref{tab:compps}).  Most strikingly we observe a clear difference in the range of best-fit $l_h/l_s$ and $\tau_p$ values for the two objects.  For the BH we find $l_h/l_s\approx12.1-12.8$, while the NS varies between $3.1-3.7$.  As the total luminosity of escaping photons must be equal to the luminosity of the input photons for this model, which is to say $l_h+l_s=l_{total}$, one can consider the $l_h/l_s$ a proxy for the amplification factor, the dichotomy in which appears to be preserved.  For the NS we find $\tau_p$ in the range $3.1-4.0$, and a range of $1.7-2.0$ for the BH.  Just as the dichotomy in spectral parameters found using \compps~ can be expressed as a different optical depth for a given $kT_e$ (figure~\ref{fig:kTetau}) and the two sources have quite different $kT_e$, the significance of the difference in $\tau_p$ is not immediately apparent.   However, $\tau_p$ and $kT_e$ seem to obey a separation between NSs and BHs in the same sense as the separation seen in figure~\ref{fig:kTetau}.  For this experiment, \xspec~ returned a range of $l_{nt}/l_h\approx 0.55-0.80$, with no clear separation between BH and NS sources.  Finally, we find $R\approx0.18-0.23$ for the BH, fully consistent with the value returned from \compps, however, for the NS we find $R\approx0.0-0.13$, significantly less than found previously.  As the treatment of the reflection is the same for both \compps~ and {\sc eqpair}  (using the {\sc ireflct} model) this suggests there is some difference in spectral shape between the two Comptonisation models.  

We emphasise that any attempt to use Comptonisation properties as a diagnostic of the nature of the compact object should first establish how the dichotomy is manifested in the specific spectral model being used.  We caution against the practice of taking any favoured Comptonisation model that might happen to feature $y$ as a free parameter, performing spectral fitting and declaring evidence in favour of a BH or NS based on the results of this paper.  Such experiments should always be prefaced by work establishing the dichotomy for that particular model using either known sources (as in this work) or by studying the effects of fitting such models to simulated \compps~ spectra based on the `typical' BH and NS spectral shapes (table~\ref{tab:compps}).

\subsection{Systematic inaccuracies in seed photon spectral parameters}
\label{subsec:seed}
On inspection the seed photon temperatures ($kT_{bb}$) and \compps~normalisations for some of the sources appear incompatible with the interpretation of the hard state in terms of a truncated disc.   In this scenario an optically thick accretion disc provides the seed photons for Comptonisation in a hot corona close to the compact object.  Truncation of the disc in the hard state is motivated by the absence of the thermal component that dominates the spectra during the high/soft state that is well-described in terms of the thermal spectrum expected from the optically thick, geometrically thin disc of \cite{1973A&A....24..337S}.  The peak temperature of this disc in the soft state is $\approx 1$~keV for BHs, and the inferred radius is consistent with the range of conceivable radii for the ISCO \citep[see review by][]{2007A&ARv..15....1D}.  From the absence of such emission it follows that the disc must either be truncated at some radius  during the hard state \citep[][]{1997ApJ...489..865E, 1997LNP...487...45G, 2001MNRAS.321..759C}, or have significant portions obscured by the Comptonising media.  However, a non-truncated disc would cool the corona too quickly; increased Comptonisation leading to a reduction in the coronal temperature but increasing the incident flux of up-scattered photons back into the disc, in turn increasing the supply of seed photons.

\begin{table*}
  \begin{tabular}{lcccccccc}
Source &  $kT_{bb}$~ & $kT_e$~ & R & y & Norm. & $kT_{in}$ & $Norm._{diskbb}$ & A \\ 
\hline
 &   keV & keV & & & & keV & &    \\
\hline
4U~1636-536 &   $0.167_{-0.008}^{+0.009}$ &   $16.7_{-2.}^{+2.}$ &   $0.22_{-0.2}^{+0.2}$ &   $0.91_{-0.03}^{+0.02}$ &   $42900._{-8500.}^{+10000.}$  & $0.24_{-0.01}^{+0.01}$ &  $2200_{-470}^{+600}$  & 3.0 \\ 
Cyg~X-1 &   $0.21_{-0.006}^{+0.01}$ &   $118._{-8.}^{+8.}$ &   $0.44_{-0.03}^{+0.03}$ &   $1.15_{-0.02}^{+0.01}$ &   $\change{99800}._{-16800.}^{+10000.}$  & $0.235_{-0.004}^{+0.004}$ & $46200_{-4500}^{+4100}$  &  7.7 \\ 
GX~339-4 &   $0.63_{-0.4}^{+0.09}$ &   $57.4_{-9.}^{+10}$ &   $0.46_{-0.08}^{+0.09}$ &   $1.32_{-0.03}^{+0.03}$ &   $284._{-134.}^{+816.}$   & ${0.196}_{-0.005}^{+0.005}$  & $69700_{-6400}^{+7000} $  & 6.7 \\ 
\end{tabular}
  \caption{As with table~\ref{tab:compps}, analysis of XMM+RXTE data, using EPIC-PN data over $0.7-9.0$ keV and RXTE data over $9.0-200$ keV. We also include the calculated value of amplification factor $A$\label{tab:caltab}. }
\end{table*}

Truncation models are disputed despite a strong theoretical justification and an ability to offer prosaic qualitative explanations for various hard state phenomena, such as the launching of jets \citep[e.g.][]{2004MNRAS.355.1105F}.  Detractors primarily argue that the broadening of emission lines by smearing due to special and general relativistic effects shows that the line emitting region, assumed to be in the disc, must be close to the compact object \citep{2006ApJ...653..525M,2010MNRAS.402..836R,2012ApJ...751...34R}. These results are in-turn disputed on the grounds of unreliable data reduction \citep[see discussions in][regarding pile up effects on line profiles]{2010MNRAS.407.2287D,2010ApJ...724.1441M},  the nature of the continuum modelling \citep[][]{2014MNRAS.437..316K} and the constituents of the line profile itself \citep[][argue that there are many ionized species present in the vicinity of the Fe $K\alpha$]{2015A&A...573A.120P} that result in an apparently broader line profile than any one individual line possesses.

For our BH sample, 3--200 keV luminosities  are typically in the  $\sim 10^{36}-1.5\times10^{38}~{\rm erg~s^{-1}}$ range, corresponding to the mass accretion rate of  $\dot{M}\sim  2\cdot 10^{-10}-3\cdot 10^{-8} {\rm M_\odot~yr^{-1}}$.  From these mass accretion rates and the lower-limit to BH masses taken from Table \ref{tab:nh}, the theory of the standard  accretion disc \citep{1973A&A....24..337S} predicts a maximal disc temperature for a non-rotating BH in the  $ kT_{max}\sim0.13-0.57$ keV range. Correcting for spectral hardening  \citep[$f\sim1.6$,][]{2011ApJ...742..122S} the observed temperatures should be  in the range $kT_{max}\sim0.2-0.9$~keV.  These values of $kT_{max}$ should be considered an upper-limit since they are calculated based on the lower mass estimate of the BHs, and we do not expect the disc to extend to the vicinity of the ISCO while in the hard state.  For example, a disc truncated at $15-100~r_g$ will have innermost temperatures reduced to $\approx 0.8-0.2$ of the theoretical maximum because $T(R)\propto R^{-0.75} (1-\sqrt{R_{*} /R} )^{0.25} $.

Across our sample of NS LMXBs  the theoretical maximum disc temperature we estimate from the  luminosities is $kT_{max}\sim0.37 - 0.63$ keV, which becomes $kT_{max}\sim 0.6 - 1.0$ keV after correcting for spectral hardening.  As for black holes, the temperature will be by a factor of a few smaller for a disc truncated at a $\sim 50-100R_g$. However, in the case of NSs a significant fraction of the seed photons may originate on the surface of the NS itself.  Taking into account that the typical luminosity enhancement factor for NS sources is $A\sim 2$ (figure~\ref{fig:fa}),  this implies an average  seed photon luminosity $\sim 0.5 \times 10^{37}~{\rm erg~s^{-1}}$ (${\rm 0.05L_{Edd}}$).  Therefore the maximum temperature of the seed photon spectra, corrected for spectral hardening  \citep[assuming correction factor of $f\sim1.6$,][]{2011ApJ...742..122S}  should  produce  a typical $ kT_{NS}\sim 1$ keV, assuming the NS radius of $\approx 15$ km.  These figures are in good agreement with an ongoing Suzaku study by Zhang et al. (in prep.).

From fitting  RXTE spectra we obtain values of  $kT_{bb}\approx0.5-1.0$~keV for BHs, and  of  $kT_{bb}\approx1.0-1.5$~keV for NSs (Table~\ref{tab:compps}). For the majority of BHs, the best fit values are a \ factor of  $\sim 1.5-5$ times larger than  the expected $kT_{max}$ for the truncated disk. The difference between measured and predicted temperatures may be much smaller in the case of NS systems, especially if the seed photon flux is dominated by the neutron star emission. 

An important limitation of PCA data is that the spectral fitting can only be performed at energies $\gtrsim3$~keV, greater than the plausible values of $kT_{bb}$, especially for BH XBs.  We utilised data from simultaneous XMM Newton observations for three sources (table~\ref{tab:xmmsource}) to scrutinise the effect of the \rxte~ bandpass on $kT_{bb}$, and other parameters.  
For each source we perform a joint analysis of the EPIC-pn spectrum between $0.7-9.0$~keV, the PCA data from $9.0-20.0$~keV and HEXTE data over $20.0-200.0$ keV.  Prior inspection of the XMM data revealed a couple of narrow lines at $\sim1.7$ keV and $\sim2.2$ keV, which are thought to be the result of incorrect compensation for the charge transfer inefficiency of the CCD \citep{2010MNRAS.407.2287D}.  We chose to ignore the data between 1.5 and 2.4 keV rather than further complicate the spectral model with additional lines.  We fixed the PCA-HEXTE calibration to the values found from our principal RXTE analysis. A clear excess can be seen at energies $\le 1$ keV, which is well-documented feature often modelled as disc emission characterised by $kT_{in}\sim0.1-0.4$ keV \citep[e.g.][]{2006ApJ...653..525M,2008MNRAS.388..753G,2010MNRAS.402..836R,2014MNRAS.437..316K}, or as an additional Comptonising component \citep{2005MNRAS.362.1435I} or as additional reflection \citep{2016arXiv160105867C}.  We chose to describe this broad excess emission as the standard geometrically-thin, optically thick accretion disc of \cite{1973A&A....24..337S} using the \xspec~model \diskbb~\citep{1984PASJ...36..741M}.  $N_H$ was allowed to vary during the fitting process because of the greater spectral resolution of XMM at energies that are especially influential to the behaviour of this parameter.  Acceptable fits were obtained for all three sources  and the recovered parameters of interest are presented in table~\ref{tab:caltab}. 
 
The  excess below 1~keV was adequately described by the \diskbb~component, and is characterised by an inner-disc temperature of $0.1-0.4$ keV, consistent with previous results \citep{2006ApJ...653..525M,2013ApJ...779...26S,2014MNRAS.437..316K}.
It should be noted that exact spectral composition of the lower energy spectral shape is the subject of continued debate \citep[for example there  are good arguments for the addition of a second Comptonised component or additional reflection,][]{2005MNRAS.362.1435I,2012PASJ...64...72S,2016arXiv160105867C}, and it is beyond the scope of the current work to reach a complete understanding of the disc geometry. Crucially, the disc seed photon temperatures are in line with expectations and other results in the literature.

As expected, there is  reduction in the recovered seed photon temperature of \compps~when one considers a wider bandpass. 
For Cyg X-1 with the  $0.7-200$ keV  luminosity of  $1.46 \times 10^{37}~\mathrm{ erg~s^{-1} }$ ($\dot{M}\approx2.6 \cdot 10^{-9}$ M$_\odot~yr^{-1}$) the maximal disc colour temperature after correcting for spectral hardening  ($f=1.6$) is $kT_{max}\approx 0.28$~keV. The measured value of $kT=0.21$ keV is consistent with the disc truncated at $ r\sim 17 r_g$ for a non-rotating BH.   In the case of GX~$339-4$, which is probably a less massive BH than Cyg~X-1, the expected maximal colour temperature of the disc spans a plausible range of $ \approx 0.3-0.8$ keV (corrected for spectral hardening). The measured value of $0.63$ keV implies a truncation could be as large as $\sim 16~R_g$.  Interestingly, for 4U~1636-536 we obtain from the broad band fit the temperature of $kT\approx 0.17$ keV, much smaller than the maximum expected for this neutron star luminosity,\footnote{note that replacing the multicolour disc seed photon spectrum with a black body spectrum further decreases the best-fit value of the seed photon temperature by 0.06~keV.} $ kT_{NS}\sim 0.80$ keV, and comparable to the value expected in the accretion disc at $r\sim 10^2 R_g$.

\begin{figure}
\begin{center}
\includegraphics[height=0.48\textwidth]{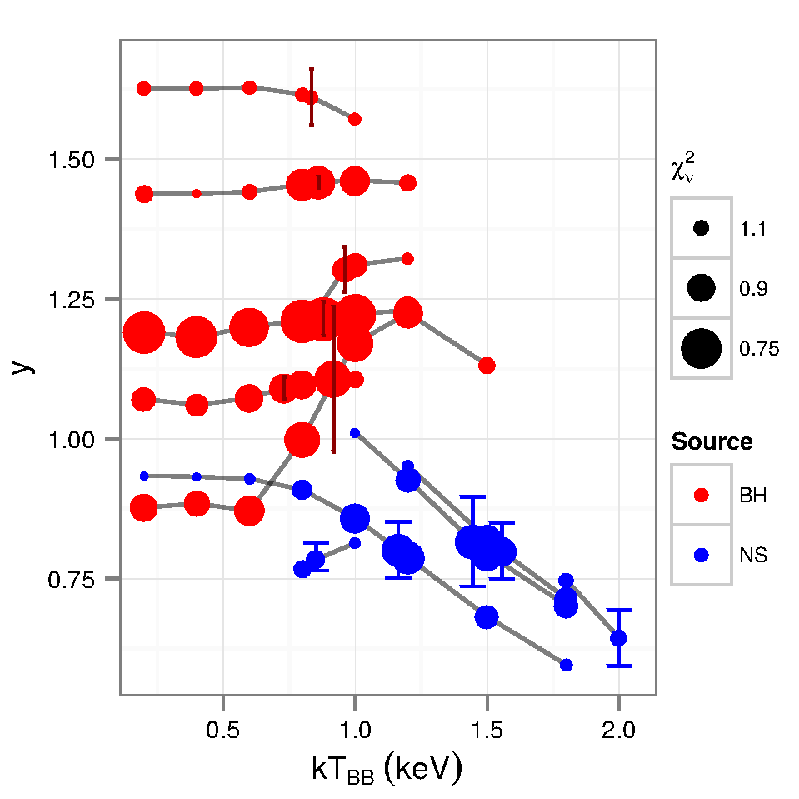}
\caption{Behaviour of $y$ with fixed values of $kT_{BB}$.  Compact object type denoted by colour, goodness of fit indicated by point size.  Points where the theoretical null hypothesis is less than $1\%$ (large chi-squared) are omitted\label{fig:ktbexperiment}.  Therefore, the lines connecting blue (red) points show the parameter region accessible to each NS (BH), with decreasing pointsize corresponding to decreasing goodness-of-fit.}
\end{center}
\end{figure}

The shift in $kT_{BB}$ between the different analyses is  accompanied by some changes in other parameters of interest, in particular in the Comptonisation parameter $y$. On the other hand,  the electron temperature $kT_{e}$ remains almost unaffected (tables \ref{tab:compps} and \ref{tab:caltab}).

In order to investigate, how  significantly the Comptonisation parameter may be affected by the bias in the seed photon temperature caused by the limited bandpass of RXTE, we conduct the following experiment. For a sub-sample of spectra (one spectrum for each source except for 4U1636-536 for which we analysed two spectra, see below), we perform simple fitting with our default spectral model, fixing $kT_{bb}$ across a grid of values.  For each $kT_{bb}$, we note the $\chi^2$, and compare it with a limit corresponding to the theoretical null hypothesis of $1\%$ for a $\chi^2$-distribution described by the number of degrees of freedom in that fit, and points with $\chi^2$ in excess of this are rejected.  This gives an indication of the space covered by potential fits that we define as being `formally acceptable'.  We present the results of this exercise in figure~\ref{fig:ktbexperiment}, which shows the behaviour of $y$-parameter with $kT_{bb}$, with the quality of the fit denoted by point size.  

From this experiment we find that BH spectra are generally compatible with a broad range of $kT_{bb}$, including low values $kT_{bb}\sim 0.2$ keV predicted by the standard accretion disc model.  While  best fits generally  prefer  $kT_{bb}\sim 1$ keV, low values of $kT_{bb}\sim 0.2$ keV   result in sufficiently small $\chi^2$, which are formally acceptable (although often outside formal $\sim 1-2\sigma$ confidence intervals). NS systems, on the other hand,  tend to favour higher values of $kT_{bb}\sim 1-1.5$ keV. For majority of NS spectra low values of $kT_{bb}\la 1$ keV result in unacceptable fit quality. This behaviour is consistent with our expectations, that the source of seed photons in BH systems is the (truncated) accretion disk, while in NS systems it is the neutron star surface, having significantly higher temperature.  BH systems are subject to a notable upward bias in the seed photon temperatures determined from fitting the 3--200 keV band  data, which   appears to be the result of the limited bandpass combined with the spectral model which is simplified and not fully adequate. However, there is virtually no  degeneracy (with the  exception of 4U~$1543-47$)  between the seed photon temperature and the $y$-parameter and electron temperature. The NS systems  are subject to the bias in the $kT_{bb}$ to much lesser extent, if at all, because of the typically higher seed photon temperatures. Although they do show some $kT_{bb}-y$ degeneracy, their seed photon temperatures are measured sufficiently well with the RXTE data. 
Therefore the  dichotomy between the BH coronal properties and those of NSs  is secure irrespective of the inaccuracies in determination of the seed spectral properties.

The observation of 4U1636-536, for which simultaneous RXTE and XMM-Newton data were analysed above, seems to differ from other NS spectra\footnote{unfortunately, this was the only NS spectrum from our sample, for which simultaneous RXTE and XMM-Newton  hard state data were available}. The  0.5-200 keV band fitting gave low $kT_{bb}\approx 0.17$ keV.  Consistent with this, in the 3-200 keV fit,  low values of $kT_{bb}\sim 0.2$ keV result in formally acceptable $\chi^2$ values.  On the other hand, the broad band data is consistent with high $kT_{bb}\sim 1$ keV, producing formally acceptable $\chi^2$ values. In either case,  the $y$-parameter is staying well below the $y\sim 1$ boundary. In interpreting these results one should bear in mind that the 0.5--200 keV and 3--200 keV data were fit with different spectral model, which in the former case included the {\tt diskbb} component.   Such behaviour is demonstrated  by only this particular observation of the source.  For comparison, we plot in figure~\ref{fig:ktbexperiment} results for  another spectrum of  4U1636-536, which in RXTE data shows a pattern  similar to other NS systems.

Any systematic bias in estimating the seed spectral properties, when considering energies $> 3$~keV, will propagate to a bias in our calculation of the the amplification factor $A$ (\S~\ref{subsec:compamp}). However, the effect of the reduced seed photon temperature on the calculated seed luminosity is  countered by a concurrent increase in model normalisation, obtained in the 0.7-200 keV band fits. Using only RXTE data Cyg~X-1 was found to have $L_{seed}=2.41\times 10^{36} \rm{erg~s^{-1}}$ (${\rm \approx0.001L_{Edd}}$), while fitting XMM+RXTE data indicates $L_{seed}=1.63\times 10^{36} \rm{erg~s^{-1}}$ (see table~\ref{tab:caltab}). Comparing the joint XMM-RXTE fitting results with those found from studying the RXTE data only, the $L_{seed}$ of 4U~1636-536 and GX~339-4 goes from $4.64\times 10^{36} \rm{erg~s^{-1}}$~to~$3.10\times 10^{36} \rm{erg~s^{-1}}$ (${\rm \approx0.025-0.017 L_{Edd}}$) and from $14.5 \times 10^{36} \rm{erg~s^{-1}}$~to~$11.5 \times 10^{36} \rm{erg~s^{-1}}$ (${\rm \approx0.07-0.06 L_{Edd}}$), respectively.  Therefore for these three sources,  the $L_{seed}$ found from fitting the $0.7-200$ keV is $\approx 0.7 L_{seed}$ found from fitting the $3-200$ keV data.   This  further demonstrates that the inaccuracy in the seed photon parameters does not affect our conclusion that the coronae in BH systems are more strongly Comptonising than those found in NS XBs. For completeness, estimating the total unabsorbed flux in the upscattered component over the full bandwidth, $0.7-200.0$~keV, leads to a marginal increase in A by 0.7, 0.05 and 0.4 for 4U~1636-536, Cyg~X-1 and GX~339-4 respectively.  The resulting values of $A$ are shown in table~\ref{tab:caltab}.

\section{Summary}
\label{sec:conc}
We have analysed hard state spectra from a sample of Galactic X-ray binaries observed by the \rxte~satellite.  The spectra were well-described by thermal Comptonisation of seed photons from an accretion disc, which we modelled using the \compps~ component in Xspec  with an additional Gaussian component to account for fluorescent emission in the vicinity of the Fe~$K\alpha$ complex.   Our most striking result is the clear dichotomy in the Compton $y-$parameter and the Compton enhancement factor $A$  between NSs and BHs (figure~\ref{fig:ampy}). The boundary is located at $y\approx 0.9$ and $A\approx 3$ with black hole systems having systematically higher  values of $y$ and $A$.  Distribution of $y$ and $A$ for neutron stars are rather narrow and strongly peaked at $y\approx 0.7-0.8$ and $A\approx 2$ ($A\sim 3$ after correction for absorption and RXTE bandpass).  Electron temperature in BH systems can occupy rather broad range of values from $\sim 30-200$ keV, while in NS their distribution  is strongly peaked at $kT_e\sim 15-25$ keV with very few spectra exceeding $kT_e\sim  50-70$ keV.  For a given optical depth the electron temperature of the Comptonising media is systematically larger in black holes than in the neutron stars (figure~\ref{fig:kTetau}).  The values of the Compton amplification factor typical for NS systems suggest that in the hard state, accreting material is losing due to Comptonisation in the corona   about $\sim 1/2-2/3$ of the total energy  it possessed upon entering the Comptonisation region. Remaining energy is released on the surface of the NS in the form of kinetic energy of the infalling material, making it a powerful source of soft radiation.  
 The smaller Compton $y-$parameter for NS LMXBs can be explained in terms of this additional supply of seed photons by the neutron star  surface  that is of course absent in the case of BHs. The observed dichotomy as a diagnostic tool has excellent potential  for distinguishing BH from NS XBs both in our galaxy and beyond.

\section*{Acknowledgments}
We acknowledge the use of the Legacy Archive for Microwave Background Data Analysis (LAMBDA), part of the High Energy Astrophysics Science Archive Center (HEASARC). HEASARC/LAMBDA is a service of the Astrophysics Science Division at the NASA Goddard Space Flight Center. MG acknowledges a partial  support by  the RFBR grant No. 15-42-02573.  MG acknowledges hospitality of the Kazan Federal University (KFU) and support by the Russian Government Program of Competitive Growth of KFU. We thank the referee for their careful reading of the manuscript and insightful suggestions that improved the quality of the paper.

\bibliography{hard}{}
\bibliographystyle{mn2e}

\bsp

\label{lastpage}

\end{document}